\documentclass[11pt]{article}

\usepackage[T1]{fontenc}
\usepackage{amsmath,amssymb}
\usepackage{graphicx}
\usepackage{caption}
\usepackage[authoryear,round]{natbib}
\usepackage{hyperref}

\title{Motion of a cylinder between parallel plates: complete Stokes resistance and pressure response}
\author{Serhat Ye\c{s}ilyurt\textsuperscript{1}\thanks{Corresponding author: \href{mailto:syesilyurt@sabanciuniv.edu}{syesilyurt@sabanciuniv.edu}}\\
and Barbaros \c{C}etin\textsuperscript{2}\\[0.5em]
\small \textsuperscript{1} Faculty of Engineering and Natural Sciences, Sabanci University\\
\small Istanbul, T\"urkiye\\
\small \textsuperscript{2} Microfluidics and Lab-on-a-chip Research Group,\\
\small Mechanical Engineering Department, Bilkent University,\\
\small 06800 Ankara, T\"urkiye}
\date{}

\begin{document}
\maketitle

\begin{abstract}
A circular cylinder moving between two parallel plates is the canonical confined problem of two-dimensional Stokes flow. Rotation and translation of the cylinder along the channel become coupled and generate a pressure difference along the channel under zero-net-flux conditions. Existing results use different methods. We develop a completed double-layer boundary integral formulation based on the channel Green function to resolve narrow gaps at arbitrary transverse positions. The channel Green kernels are derived from a Fourier-transformed biharmonic problem, permitting explicit treatment of the free-space singularity, the slowly decaying nearest-wall contribution, and the small-wavenumber limit for accurate kernel evaluation under strong confinement. The boundary integral equation is discretised by a spectrally accurate Nystr\"om scheme on a graded parametrisation, and the pressure jump follows from the zero-wavenumber limit of the streamwise momentum balance. A single linear system yields the full resistance matrix and pressure-jump coefficients for an eccentric cylinder; an additional right-hand side treats imposed net flux. For nonzero coefficients, comparisons with independent COMSOL finite-element simulations show relative differences below \(0.1\%\) for the resistance coefficients and at most \(0.0122\%\) for the pressure-response coefficients and excess hydraulic resistance. Reciprocity, symmetry constraints, and lubrication asymptotics provide additional checks on the numerical results. We track the coupling between rotation and translation and the pressure generated by the cylinder from the channel scale into the lubrication regime, and determine the pressure--flow relation, driving torque and efficiency of the rotating-cylinder pump for asymmetric clearances.
\end{abstract}

\section{Introduction}
The motion of particles near solid boundaries is fundamental to sedimentation in confined suspensions, particle and cell transport through microchannels, and the operation of microscale pumps, rotors and robotic actuators. A cylinder between parallel plates provides a canonical two-dimensional model for these systems. It has been used to study the settling and rotation of elongated particles in narrow channels \citep{Dvinsky1987a}, their mobility in pressure-driven microflows \citep{Dvinsky1987b,Jeong2014b}, viscous pumping by an eccentrically rotating element \citep{Sen1996,Day2000}, and the motion of rotating filaments in tightly confined biological spaces \citep{Yang2011,Yang2013}. In these settings, the force and torque on the cylinder, the coupling between translation and rotation, and the pressure response as a function of  flow rate along  the channel are the principal hydrodynamic quantities.

The confined problem differs fundamentally from motion of a cylinder near a plane wall in an unbounded fluid.  For the infinite-channel resistance problem considered here, fluid displaced by the moving cylinder must return through the gaps between the cylinder and the plates. This return flow generates a finite end-to-end pressure difference and, when the cylinder is eccentric, couples parallel translation and rotation: a rotating cylinder experiences a streamwise force, while a translating cylinder experiences a torque \citep{Champmartin2007b,Yang2013}. The problem becomes especially demanding as the minimum clearance $h$ decreases. The local lubrication region has a streamwise extent of order $(ah)^{1/2}$, whereas the return flow and pressure field remain controlled by the channel scale. A calculation must consequently resolve the singular gap stresses and the global flux constraint simultaneously.

Classical analytical results cover several important limiting cases. For a cylinder on the channel centreline, symmetry eliminates translation--rotation coupling and permits separable solutions. Series expansions were used to determine the drag on a translating cylinder \citep{Faxen1946,Takaisi1955,Harper1967}, while \citet{Hellou1992} calculated the torque and flow structure generated by a rotating cylinder. Near a single plane wall, \citet{Jeffrey1981} obtained the exact resistance coefficients for translation and rotation. In that strictly two-dimensional single-wall geometry, rotation produces torque but no net force, and parallel translation produces no torque. The leading resistances diverge as $(h/a)^{-3/2}$ for translation normal to the wall and as $(h/a)^{-1/2}$ for parallel translation and rotation. These single-wall results determine the dominant local singularities when one channel gap closes, but they contain neither the channel-scale return flow nor the associated pressure jump and coupling introduced by the second plate.

When both gaps are small relative to the cylinder radius, lubrication theory incorporates the global flux constraint by coupling the flows through the two clearances. \citet{Day2000} analysed an eccentrically rotating cylinder as a viscous pump, while \citet{Yang2013} considered combined rotation and parallel translation. Finite-volume calculations further demonstrated that, for a cylinder nearly blocking the channel, the parallel drag and pressure drop can grow as the clearance to the power $-5/2$, whereas the rotational torque grows as the clearance to the power $-1/2$ \citep{Champmartin2007a,Champmartin2007b}. These asymptotic descriptions apply when both gaps are narrow, but do not generally describe the asymmetric regime in which one clearance is small while the other remains comparable to the cylinder radius. This regime is particularly important because the nearest wall controls the singular local resistance, while the opposite wall influences the return flow, pressure response, and translation--rotation coupling.

Following the initial proposal of \citet{Sen1996}, rotating-cylinder pumps were investigated numerically at small but finite Reynolds numbers \citep{Sharatchandra1997,DeCourtye1998}. In the Stokes limit, \citet{Day2000} derived lubrication relations among the rotation rate, driving torque, flow rate, and pressure difference, and supplemented these with numerical boundary-integral solutions of the full Stokes equations. Their method employed the free-space Stokeslet and stresslet in a direct boundary-integral formulation, requiring discretisation of the cylinder surface, both channel walls, and artificial upstream and downstream boundaries. The unknown surface tractions were approximated using piecewise-linear boundary elements with local refinement near the narrow clearances, while the excess pressure difference was extracted from the wall tractions after subtracting the Poiseuille contribution. Their results established the fundamental pumping mechanism, quantified deviations from lubrication theory, and examined the influence of geometry on pumping performance.

Eccentric cylinders at finite clearance have consequently been treated mainly by numerical methods. \citet{Dvinsky1987a} solved the vorticity--streamfunction equations on boundary-conforming finite-difference grids and reported coefficients for all three rigid-body modes together with the pressure drop generated by cylinder motion. The companion study used these coefficients to determine the motion of a force- and torque-free cylinder in imposed channel flows \citep{Dvinsky1987b}. Finite-volume and finite-element calculations subsequently examined confined translation, rotation, near-contact scaling, and reciprocal force--torque coupling \citep{Champmartin2007a,Champmartin2007b,Yang2013}. Papkovich--Fadle eigenfunction expansions have also been used to calculate forces, moments, and pressure changes for cylinders in quiescent and pressure-driven channels at arbitrary transverse positions \citep{Jeong2014a,Jeong2014b}. Together these studies cover much of the finite-clearance response, but the available values are dispersed among different motion modes and conventions, and verified accuracy in the narrow-gap eccentric regime remains limited. Volume discretisations require a truncated channel and strong local mesh refinement, while the resolution of a narrow lubrication region by channel-scale eigenfunctions has not been demonstrated.

Boundary-integral methods offer a natural alternative because the Stokes equations are satisfied in the fluid domain and only boundaries need to be represented. In a direct formulation, the unknown is the physical surface traction and the boundary equation is of the first kind. \c{C}etin and co-workers developed direct boundary-element formulations for tracking multiple particles and multi-domain particulate flow in microchannels, and extended this approach to isogeometric simulations of confined droplets \citep{Karakaya2015,Topuz2021,Gumus2025}. For the present narrow-gap resistance problem, however, first-kind systems become increasingly ill-conditioned under refinement \citep{Karrila1989,PhanThien1992}. Moreover, when the free-space fundamental solution is used, the plates and artificial channel ends must be discretised together with the cylinder, as in the pump calculation of \citet{Day2000}. These requirements are especially costly when a narrow gap demands strong local refinement.

The indirect completed double-layer formulation avoids both limitations. A pure double layer carries neither net force nor net torque and has a null space associated with rigid-body motions. Supplementing it with interior Stokeslet and rotlet singularities and appropriate constraints produces a uniquely solvable second-kind equation, while the singularity strengths give the resultant force and torque directly \citep{Power1987,Karrila1989,Power1993a}. Completed double-layer formulations were developed for particles near a single plane wall \citep{PhanThien1992,Power1993b,Kohr1998}. Related wall-adapted Green functions and boundary-integral representations have also been used for three-dimensional motion between parallel plates, including the fundamental solution of \citet{Liron1976} and calculations for confined spheres \citep{Staben2003,Bhattacharya2005}. These studies establish the principal advantages of combining completion singularities with Green functions tailored to the confining geometry.

For a two-dimensional channel, the wall-bounded Green function can be constructed using Fourier transformation in the streamwise direction \citep{Pozrikidis1987,Davis1993,Crowdy2013}. \citet{Kohr2000} developed a formulation combining single- and double-layer potentials, while \citet{Kohr2005} established a completed double-layer integral equation of the second kind for smooth obstacles between parallel plates, using the previously derived channel Green function of \citet{Pozrikidis1987}. Since this Green function satisfies the no-slip conditions on both plates, only the obstacle boundary needs to be discretised. \citet{Kohr2005} also established existence and uniqueness of the formulation, which accommodates general rigid-body motion, including rotation. However, the numerical implementation employed piecewise-constant boundary elements and was demonstrated only for the drag on a circular cylinder translating parallel or perpendicular to the walls. Rotational resistance, translation--rotation coupling, pressure differences across the obstacle, and systematic narrow-gap calculations were not reported.

The present study builds upon the completed double-layer framework of \citet{Kohr2005}, but develops a different approach to the construction and numerical evaluation of the channel Green kernels. Rather than directly employing the Green tensor of \citet{Pozrikidis1987}, we construct the wall-bounded fundamental solution from a scalar biharmonic Green problem obtained by Fourier transformation of the streamfunction equation. The channel Stokeslet, its associated stress kernel, and the rotlet completion are then obtained through derivatives of this scalar Green function. The resulting representation permits explicit treatment of the free-space singularity, the slowly decaying near-wall contributions, and the small-wavenumber limit, enabling accurate kernel evaluation under strong confinement. In contrast to the direct boundary-integral approach of \citet{Day2000}, which employed free-space singularities and discretised the complete boundary of a finite computational domain, the present completed double-layer formulation incorporates the no-slip conditions on both plates analytically, requiring discretisation only of the cylinder boundary and eliminating artificial inlet and outlet boundaries. The pressure difference is extracted directly from the zero-wavenumber contribution, without evaluating pressures in the distant flow field, while force and torque follow from the completion strengths. Combined with a graded Nystr\"om discretisation and independent finite-element validation, this approach yields the complete resistance matrix and pressure-response coefficients, extending the analysis beyond translational drag and fixed-axis pumping to coupled rigid-body motion. The resulting unified formulation enables quantitative investigation of force-free backward sliding, confinement-dependent hydrodynamic resistance, and the pressure--flow, torque and efficiency characteristics of rotating-cylinder pumps, including configurations with strongly asymmetric clearances.

This formulation places rigid-body resistance, pressure generation, freely suspended motion, and viscous pumping within the same linear framework. An additional right-hand side for a stationary cylinder in imposed Poiseuille flow supplies the excess hydraulic resistance, while the reciprocal theorem connects the pressure generated by rotation to the torque generated by through-flow. The resulting coefficients quantify how the second plate modifies the near-wall resistance and translation--rotation coupling, and provide the pressure--flux relation, torque, and force required to characterise an eccentrically rotating pump beyond the range of two-gap lubrication theory.

\section{Methodology}
\label{sec:methodology}
\subsection{Geometry and governing equations}
\label{sec:geometry-governing}

\begin{figure}[t]
\centering
\includegraphics[width=0.8\textwidth]{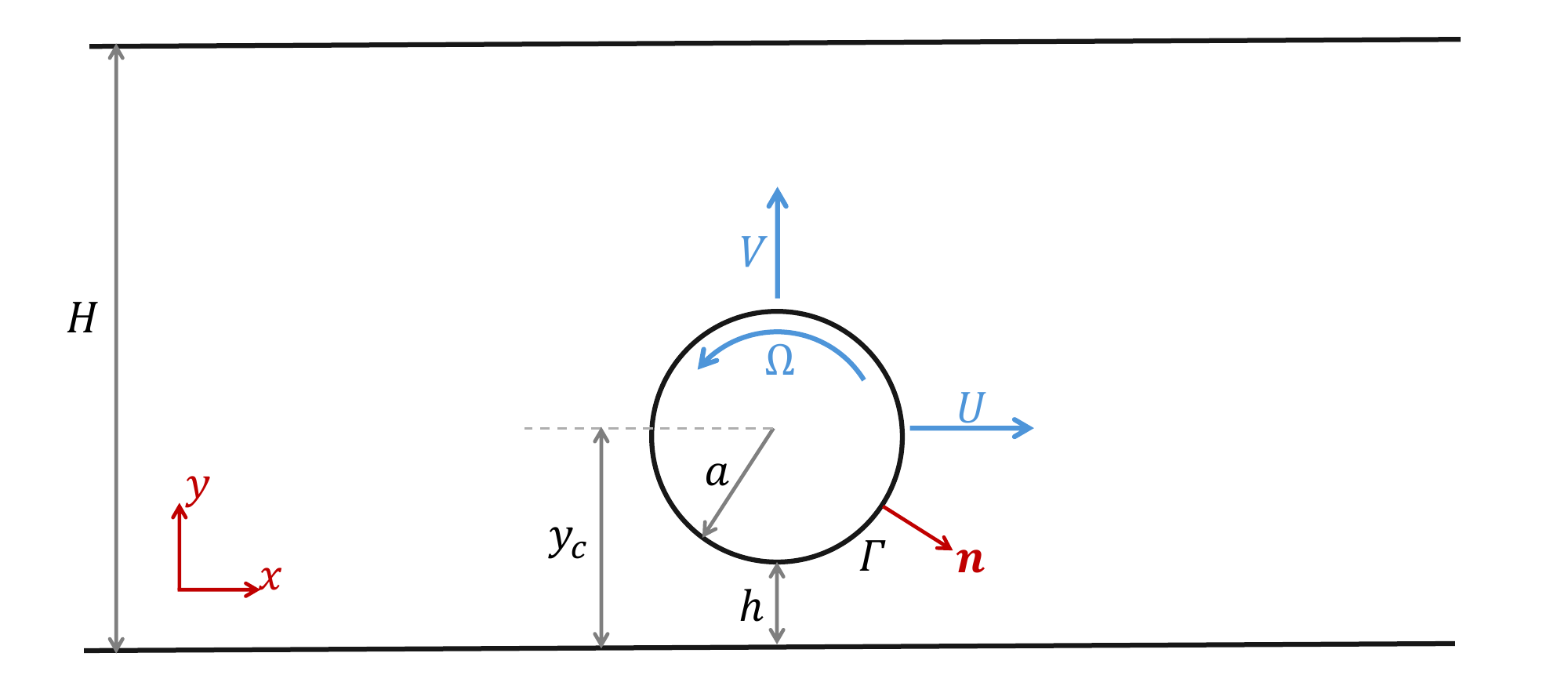}
\caption{Geometry and prescribed motion of a circular cylinder between stationary parallel plates. The cylinder has radius \(a\), centre height \(y_c\), and lower-wall clearance \(h\) in a channel of height \(H\); \(U\), \(V\), and \(\Omega\) denote its parallel velocity, normal velocity, and angular velocity.}
\label{fig:geometry}
\end{figure}

We consider the two-dimensional motion of an infinitely long circular cylinder of radius \(a\) in an incompressible Newtonian fluid of dynamic viscosity \(\mu\), confined between two stationary parallel plates, as shown in figure~\ref{fig:geometry}. The plates are located at \(y=0\) and \(y=H\), and the channel extends to infinity in the \(x\)-direction. The cylinder axis is aligned with the channel's spanwise direction, and its instantaneous centre is chosen as \((0,y_c)\), where \(a<y_c<H-a\). The minimum gap between the cylinder and the lower plate is
\begin{equation}
h=y_c-a,
\end{equation}
while the upper gap is \(H-2a-h\). The geometry is therefore specified by the two dimensionless parameters \(a/H\) and \(h/a\).

The cylinder translates with velocity \(U\boldsymbol e_x+V\boldsymbol e_y\) and rotates with angular velocity \(\Omega\boldsymbol e_z\), with positive \(\Omega\) corresponding to counterclockwise rotation in the \((x,y)\)-plane. Fluid inertia is neglected, and the flow is determined from the instantaneous cylinder position and velocity. The velocity field \(\boldsymbol u=(u,v)\) and pressure \(p\) satisfy the Stokes equations in the fluid domain,
\begin{equation}
-\boldsymbol\nabla p+\mu\nabla^2\boldsymbol u=\boldsymbol 0,
\qquad
\boldsymbol\nabla\boldsymbol{\cdot}\boldsymbol u=0.
\end{equation}

The cylinder boundary, denoted by \(\Gamma\), is parametrised by
\begin{equation}
\boldsymbol x_b(\theta)
=
a\cos\theta\,\boldsymbol e_x
+
\bigl(y_c+a\sin\theta\bigr)\boldsymbol e_y,
\qquad 0\leq\theta<2\pi.
\end{equation}
We define the unit normal directed from the cylinder into the fluid and the counterclockwise unit tangent as
\begin{equation}
\label{eq:normal-tangent}
\boldsymbol n
=
\cos\theta\,\boldsymbol e_x+\sin\theta\,\boldsymbol e_y,
\qquad
\boldsymbol t
=
-\sin\theta\,\boldsymbol e_x+\cos\theta\,\boldsymbol e_y.
\end{equation}

Thus, \(\boldsymbol n\) is opposite to the outward normal of the fluid domain on \(\Gamma\). The no-slip condition on the cylinder is
\begin{equation}
\label{eq:cylinder-velocity}
\boldsymbol u\bigl(\boldsymbol x_b(\theta)\bigr)
=
U\boldsymbol e_x+V\boldsymbol e_y+a\Omega\boldsymbol t
=
\bigl(U-a\Omega\sin\theta\bigr)\boldsymbol e_x
+
\bigl(V+a\Omega\cos\theta\bigr)\boldsymbol e_y.
\end{equation}
The stationary plates impose
\begin{equation}
\boldsymbol u(x,0)=\boldsymbol u(x,H)=\boldsymbol 0.
\end{equation}

There is no imposed through-flow. Far from the cylinder, the fluid is at rest and the pressure approaches finite constants,
\begin{equation}
\boldsymbol u\rightarrow\boldsymbol 0,
\qquad
p\rightarrow p_\pm
\quad\text{as}\quad x\rightarrow\pm\infty.
\end{equation}
These conditions select zero volume flux through channel cross-sections sufficiently far from the cylinder. The two limiting pressures need not be equal: cylinder motion can generate a finite pressure difference,
\begin{equation}
\label{eq:pressure-definition}
\Delta p=p_--p_+.
\end{equation}
The pressure difference is an output of the solution, while an arbitrary additive pressure constant may be fixed independently.

\subsection{Hydrodynamic forces and resistance coefficients}
\label{sec:hydrodynamic-loads}

The hydrodynamic force and torque on the cylinder are obtained from the fluid stress,
\begin{equation}
\boldsymbol{\sigma}
=
-p\boldsymbol I
+\mu\left[
\boldsymbol\nabla\boldsymbol u+
(\boldsymbol\nabla\boldsymbol u)^{\mathsf T}
\right].
\end{equation}
Using the normal \(\boldsymbol n\) directed from the cylinder into the fluid, as defined in \eqref{eq:normal-tangent}, the force and axial torque per unit cylinder length are
\begin{equation}
\boldsymbol F
=
\int_\Gamma \boldsymbol{\sigma}\boldsymbol n\,\mathrm ds,
\qquad
T
=
\int_\Gamma
\left[
(\boldsymbol x-\boldsymbol x_c)
\times(\boldsymbol{\sigma}\boldsymbol n)
\right]\boldsymbol{\cdot}\boldsymbol e_z\,\mathrm ds,
\end{equation}
where \(\boldsymbol x_c=(0,y_c)\). Throughout, these quantities denote the loads exerted by the fluid on the cylinder. 

Linearity of the Stokes equations implies a linear relation between the hydrodynamic loads and the cylinder velocities. Reflection symmetry about \(x=0\) decouples normal translation from parallel translation and rotation. The resistance relation therefore takes the form
\begin{equation}
\label{eq:dimensional-resistance}
\begin{pmatrix}
F_x\\[2pt]F_y\\[2pt]T
\end{pmatrix}
=
\begin{pmatrix}
R^{F_xU} & 0 & R^{F_x\Omega}\\
0 & R^{F_yV} & 0\\
R^{TU} & 0 & R^{T\Omega}
\end{pmatrix}
\begin{pmatrix}
U\\[2pt]V\\[2pt]\Omega
\end{pmatrix}.
\end{equation}
Following the notation of \citet{Teng2022}, the superscripts identify the load and the motion responsible for it. Thus, \(R^{F_x\Omega}\) is the force generated by rotation, while \(R^{TU}\) is the torque generated by parallel translation. The Lorentz reciprocal theorem gives
\begin{equation}
R^{F_x\Omega}=R^{TU}.
\end{equation}


Since the force and torque are expressed per unit length, we define dimensionless resistance coefficients by
\begin{equation}
\begin{aligned}
\widehat R^{F_xU}&=\frac{R^{F_xU}}{\mu},
&
\widehat R^{F_yV}&=\frac{R^{F_yV}}{\mu},
\\
\widehat R^{F_x\Omega}&=\frac{R^{F_x\Omega}}{\mu a},
&
\widehat R^{TU}&=\frac{R^{TU}}{\mu a},
&
\widehat R^{T\Omega}&=\frac{R^{T\Omega}}{\mu a^2}.
\end{aligned}
\end{equation}
The normalised resistance relation is then
\begin{equation}
\label{eq:dimensionless-resistance}
\begin{pmatrix}
F_x\\[2pt]F_y\\[2pt]T/a
\end{pmatrix}
=
\mu
\begin{pmatrix}
\widehat R^{F_xU} & 0 & \widehat R^{F_x\Omega}\\
0 & \widehat R^{F_yV} & 0\\
\widehat R^{TU} & 0 & \widehat R^{T\Omega}
\end{pmatrix}
\begin{pmatrix}
U\\[2pt]V\\[2pt]a\Omega
\end{pmatrix}.
\end{equation}
All entries of the dimensionless matrix depend only on the geometry, specified by \(a/H\) and \(h/a\). The scaling preserves the reciprocal equality of the two coupling coefficients.

\subsection{Translation--rotation coupling}
\label{sec:translation-rotation-coupling}

Due to symmetry with respect to the centreline, a cylinder rotating at the centreline experiences no parallel hydrodynamic force. Away from the centreline, symmetry permits a nonzero coupling. For a cylinder driven by an external torque and subject to no external force, force balance requires $F_x=F_y=0$. Equation~\eqref{eq:dimensional-resistance} then gives $V=0$, while the parallel velocity satisfies
\begin{equation}
\frac{U}{a\Omega}
=
-\frac{\widehat R^{F_x\Omega}}{\widehat R^{F_xU}},
\qquad \Omega\ne0.
\end{equation}
The dimensionless translation rate is therefore determined entirely by the geometry. It is independent of viscosity and, within the Stokes approximation, of the magnitude of the imposed rotation rate. Its sign specifies the direction of translation for a given sense of rotation.

The corresponding torque is
\begin{equation}
T
=
\mu a^2
\left[
\widehat R^{T\Omega}
-
\frac{\bigl(\widehat R^{F_x\Omega}\bigr)^2}
{\widehat R^{F_xU}}
\right]\Omega.
\end{equation}
This expression accounts for the change in rotational resistance when the cylinder is allowed to translate freely.

\subsection{Pressure difference generated by cylinder motion}

The pressure difference defined in \eqref{eq:pressure-definition} is also linear in the prescribed velocities. We introduce dimensionless pressure-response coefficients through
\begin{equation}
\label{eq:pressure-response}
\Delta p
=
\frac{\mu}{H}
\left(
P^U U+P^V V+P^\Omega a\Omega
\right).
\end{equation}
Like the dimensionless resistance coefficients, \(P^U\), \(P^V\) and \(P^\Omega\) depend only on the cylinder geometry and position.

Reflection about \(x=0\) leaves the normal-translation problem unchanged but interchanges the two channel ends. Hence normal translation produces no far-field pressure difference, and \(P^V=0\). Reflection about the channel centreline further gives
\begin{equation}
P^U(H-y_c)=P^U(y_c),
\qquad
P^\Omega(H-y_c)=-P^\Omega(y_c).
\end{equation}
Consequently, rotation produces no pressure difference when the cylinder lies at the centreline, whereas parallel translation may do so. These relations, together with resistance reciprocity and the vanishing coupling at the centreline, provide independent checks on the solution.

\subsection{Operation as a viscous pump}
\label{sec:viscous-pump}

The same geometry can operate as a viscous pump when the cylinder rotates about a fixed axis, with \(U=V=0\) \citep{Sen1996,Day2000}. The zero-flux rotational solution considered above describes the shut-off condition. To describe operation at nonzero flux, we retain the infinite-channel geometry and introduce a fully developed Poiseuille flow far from the cylinder.

Let \(Q\) denote the volume flux per unit span, positive in the \(x\)-direction. The corresponding undisturbed Poiseuille velocity and pressure are
\begin{equation}
\label{eq:pump-far-field}
\boldsymbol u_P(y)
=
\frac{6Q}{H^3}y(H-y)\boldsymbol e_x,
\qquad
p_P(x)
=
-\frac{12\mu Q}{H^3}x.
\end{equation}
We decompose the total solution into this known background and a disturbance,
\begin{equation}
\label{eq:pump-decomposition}
\boldsymbol u=\boldsymbol u_P+\boldsymbol u',
\qquad
p=p_P+p'.
\end{equation}
For \(Q\ne0\), the total pressure varies linearly at large \(\lvert x\rvert\) and is therefore unbounded in the infinite channel. The bounded-pressure condition of \S\ref{sec:geometry-governing} now applies to the disturbance pressure \(p'\), rather than to the total pressure.

Because the Poiseuille field satisfies the Stokes equations and the stationary-wall conditions, the disturbance satisfies the same homogeneous Stokes equations as the original problem, with
\begin{equation}
\label{eq:pump-disturbance-conditions}
\begin{aligned}
\boldsymbol u'(\boldsymbol x_b)
&=
a\Omega\boldsymbol t(\boldsymbol x_b)
-\boldsymbol u_P(y_b),
&&\boldsymbol x_b\in\Gamma,
\\
\boldsymbol u'&=\boldsymbol0,
&&y=0,H,
\\
\boldsymbol u'&\rightarrow\boldsymbol0,
\qquad p'\rightarrow p'_\pm,
&&x\rightarrow\pm\infty.
\end{aligned}
\end{equation}
where \(y_b=y_c+a\sin\theta\) is the vertical coordinate of \(\boldsymbol x_b\). Thus the disturbance has zero far-field flux and finite limiting pressures. It can be calculated using the same channel Green function and completed boundary-integral formulation developed below; only the prescribed velocity on the cylinder changes.

The finite difference
\begin{equation}
\label{eq:excess-pressure-definition}
\Delta p_{\mathrm{ex}}(\Omega,Q)=p'_- -p'_+,
\end{equation}
defines the excess pressure drop associated with the cylinder as a function of the imposed angular velocity and flux. Equivalently, it is the difference between the constant offsets of the upstream and downstream pressure profiles after subtracting their common Poiseuille gradient. At \(Q=0\), it reduces to the pressure difference \(\Delta p\) defined in \eqref{eq:pressure-definition}.

By linearity, superposing the zero-flux rotational solution and the stationary-cylinder through-flow solution gives
\begin{equation}
\label{eq:pump-pressure-flux}
\Delta p_{\mathrm{ex}}(\Omega,Q)
=
\frac{\mu a\Omega}{H}P^\Omega
+
\frac{\mu Q}{H^2}C,
\end{equation}
where \(P^\Omega\) is the rotational pressure coefficient introduced in \eqref{eq:pressure-response}, and \(C\) is the dimensionless excess hydraulic resistance. This relation characterises the pressure response of the cylinder after subtraction of the empty-channel pressure drop.

To calculate \(C\), we use the special case \(\Omega=0\): the cylinder is held stationary and a nonzero flux \(Q\) is prescribed. The disturbance boundary condition is then \(\boldsymbol u'=-\boldsymbol u_P\) on \(\Gamma\), so that the total velocity vanishes on the cylinder. Equation~\eqref{eq:pump-pressure-flux} reduces to \(\Delta p_{\mathrm{ex}}(0,Q)=\mu Q C/H^2\), and hence
\begin{equation}
\label{eq:excess-hydraulic-resistance}
C
=
\frac{H^2}{\mu Q}\,
\Delta p_{\mathrm{ex}}(0,Q).
\end{equation}
Linearity makes \(C\) independent of the magnitude of \(Q\); it depends only on the geometry and is positive with the present pressure-drop convention. The dimensional excess hydraulic resistance is \(\mu C/H^2\). This coefficient requires one additional boundary-integral solve beyond the three rigid-body motions used to determine the resistance matrix. The numerical calculation is described in \S\ref{sec:pump-calculation}, and the pressure functional is derived in appendix~\ref{app:pressure}.

The torque response follows from the same superposition. Applying the Lorentz reciprocal theorem to the zero-flux rotational solution and the stationary-cylinder through-flow solution gives
\begin{equation}
\label{eq:pump-reciprocity}
-T^{(Q)}\Omega
=
\Delta p_{\mathrm{ex}}(\Omega,0)\,Q,
\end{equation}
where \(T^{(Q)}\) is the hydrodynamic torque on the stationary cylinder in the through-flow problem. With the hydrodynamic-load convention of \S\ref{sec:hydrodynamic-loads}, the total torque is therefore
\begin{equation}
\label{eq:pump-torque}
T =
\mu a^2\widehat R^{T\Omega}\Omega
-
\frac{\mu aQ}{H}P^\Omega.
\end{equation}
Thus \(P^\Omega\) determines both cross responses, while \(C\) supplies the additional hydraulic resistance. Equation~\eqref{eq:pump-reciprocity} also provides a direct numerical reciprocity check between the rotational and through-flow solutions; its derivation is given in appendix~\ref{app:pressure}.

For nonzero flux, a pressure difference measured over a finite length also includes the Poiseuille contribution. Consider two sections \(x_-\) and \(x_+\), sufficiently far from the cylinder that the flow is fully developed, separated by \(L=x_+-x_-\). Their total pressure difference is
\begin{equation}
\label{eq:pump-total-pressure}
\begin{aligned}
\Delta p_L
&\equiv p(x_-)-p(x_+)
\\
&\simeq
\frac{12\mu QL}{H^3}
+\Delta p_{\mathrm{ex}}(\Omega,Q)
\\
&=
\frac{\mu a\Omega}{H}P^\Omega
+
\frac{\mu Q}{H^2}
\left(C+\frac{12L}{H}\right).
\end{aligned}
\end{equation}
The approximation neglects the remaining decaying disturbances at the measurement sections. This expression evaluates the pressure difference between sections of the infinite-channel solution. It also approximates a finite channel when the end conditions reproduce the developed flow and the cylinder is sufficiently remote from both ends; finite-end disturbances and entrance losses are not included.

With this pressure-drop convention, the pressure rise delivered in the positive \(x\)-direction is \(-\Delta p_L\). Equal pressures at the two sections give the free-delivery flux
\begin{equation}
\label{eq:pump-free-delivery}
Q_{\mathrm{free}}
\simeq
-\frac{a\Omega H\,P^\Omega}
{C+12L/H}.
\end{equation}
Setting only \(\Delta p_{\mathrm{ex}}(\Omega,Q)=0\) in \eqref{eq:pump-pressure-flux} would instead give \(Q=-a\Omega H P^\Omega/C\), which excludes the surrounding channel resistance. Equation~\eqref{eq:pump-free-delivery} shows that the operating flux depends on both the local cylinder response and the measurement length, and tends to zero as \(L\rightarrow\infty\) at fixed \(\Omega\). At the centreline, \(P^\Omega=0\), so rotation alone produces neither a shut-off pressure difference nor a net flux between equal-pressure sections.

\section{Boundary integral solution}
\subsection{Green function for the channel}

We construct a Green function that satisfies the boundary conditions on both plates, allowing the subsequent boundary-integral formulation to involve only the cylinder surface. Introducing the streamfunction \(\psi\) through
\begin{equation}
\label{eq:streamfunction-definition}
u=\frac{\partial\psi}{\partial y},
\qquad
v=-\frac{\partial\psi}{\partial x},
\end{equation}
satisfies incompressibility identically. Taking the curl of the Stokes momentum equation eliminates the pressure. In regions without singular forcing, the streamfunction satisfies
\begin{equation}
\nabla^4\psi=0.
\end{equation}

The channel is unbounded in the \(x\)-direction, which we treat using the Fourier transform
\begin{equation}
\widehat g(k,y)
=
\int_{-\infty}^{\infty}
g(x,y)\,\mathrm e^{-ikx}\,\mathrm dx,
\qquad
g(x,y)
=
\frac{1}{2\pi}
\int_{-\infty}^{\infty}
\widehat g(k,y)\,\mathrm e^{ikx}\,\mathrm dk.
\end{equation}
Here \(k\) is the streamwise wavenumber. For a source at height \(\eta\), the scalar transformed Green function \(G_k(y,\eta)\) is defined by
\begin{equation}
\label{eq:green-equation}
\left(\frac{\partial^2}{\partial y^2}-k^2\right)^2
G_k(y,\eta)
=
\delta(y-\eta),
\qquad 0<y,\eta<H,
\end{equation}
subject to
\begin{equation}
G_k(0,\eta)=G_k(H,\eta)=0,
\qquad
\partial_yG_k(0,\eta)=\partial_yG_k(H,\eta)=0.
\end{equation}
These conditions incorporate stationary no-slip walls and the zero-through-flow condition. In streamfunction form, zero through-flow permits the same streamfunction value on both plates; the arbitrary common value is set to zero.

For \(k\ne0\), the homogeneous solutions of \eqref{eq:green-equation} are linear combinations of
\(\mathrm e^{|k|y}\), \(y\mathrm e^{|k|y}\),
\(\mathrm e^{-|k|y}\) and \(y\mathrm e^{-|k|y}\).
The solutions below and above the source are matched by requiring
\begin{equation}
[G_k]_\eta
=
[\partial_yG_k]_\eta
=
[\partial_y^2G_k]_\eta
=
0,
\qquad
[\partial_y^3G_k]_\eta=1,
\end{equation}
where \([g]_\eta=g(\eta^+)-g(\eta^-)\). The unit jump follows by integrating \eqref{eq:green-equation} across the source. Together, the wall and matching conditions determine \(G_k\) uniquely. Self-adjointness of the scalar operator gives
\begin{equation}
G_k(y,\eta)=G_k(\eta,y),
\end{equation}
and \(G_{-k}=G_k\), since the defining problem depends only on \(k^2\). The explicit construction is given in appendix~\ref{app:green}. At \(k=0\), the corresponding solutions are piecewise cubic polynomials, yielding a regular zero-wavenumber limit.

The scalar Green function generates the velocity response to a point force. Consider a force per unit cylinder length \(\boldsymbol f=(f_x,f_y)\) applied to the fluid at \((0,\eta)\). With the force entering the momentum equation as \(+\boldsymbol f\,\delta(x)\delta(y-\eta)\), the transformed streamfunction is
\begin{equation}
\widehat\psi(k,y)
=
\frac{1}{\mu}
\left[
f_x\,\partial_\eta G_k(y,\eta)
+
ikf_y\,G_k(y,\eta)
\right].
\end{equation}
Applying the velocity operators in \eqref{eq:streamfunction-definition} therefore gives
\(\widehat{\boldsymbol u}=\widehat{\boldsymbol S}\boldsymbol f\), where
\begin{equation}
\widehat{\boldsymbol S}(k;y,\eta)
=
\frac{1}{\mu}
\begin{pmatrix}
\partial_y\partial_\eta G_k
&
ik\,\partial_yG_k
\\[2pt]
-ik\,\partial_\eta G_k
&
k^2G_k
\end{pmatrix}.
\end{equation}
Here \(\partial_y\) differentiates with respect to the observation height, whereas \(\partial_\eta\) differentiates with respect to the source height.

For a source at \(\boldsymbol x'=(\xi,\eta)\) and an observation point \(\boldsymbol x=(x,y)\), the channel Stokeslet is obtained by inverse transformation:
\begin{equation}
\label{eq:channel-stokeslet}
\boldsymbol S(\boldsymbol x,\boldsymbol x')
=
\frac{1}{2\pi}
\int_{-\infty}^{\infty}
\widehat{\boldsymbol S}(k;y,\eta)
\,\mathrm e^{ik(x-\xi)}\,\mathrm dk.
\end{equation}
The velocity induced by the point force is then
\(\boldsymbol u(\boldsymbol x)=
\boldsymbol S(\boldsymbol x,\boldsymbol x')\boldsymbol f\).
Every column of \(\boldsymbol S\) satisfies no slip at both plates.

\subsection{Completed double-layer representation}

The channel Green function satisfies the boundary conditions on both plates. We now construct a velocity field from this Green function and determine its unknown coefficients by imposing the prescribed velocity on the cylinder. The representation has two parts: a double-layer distribution over the cylinder boundary and a completion field that accounts for the net force and torque.

\paragraph{The double-layer field.} Let \(\boldsymbol\phi(\boldsymbol x')\) be an unknown two-component density defined on the cylinder boundary \(\Gamma\). Its associated double-layer velocity is
\begin{equation}
\label{eq:double-layer-field}
\boldsymbol u^{D}(\boldsymbol x)
=
\int_\Gamma
\boldsymbol K(\boldsymbol x,\boldsymbol x')
\,\boldsymbol\phi(\boldsymbol x')\,\mathrm ds',
\qquad \boldsymbol x\notin\Gamma.
\end{equation}
Here \(\boldsymbol x\) is the observation point, \(\boldsymbol x'\) is the integration point on the cylinder, and \(\mathrm ds'\) is the boundary arclength element. The \(2\times2\) kernel \(\boldsymbol K\) includes the normal \(\boldsymbol n(\boldsymbol x')\), directed from the cylinder into the fluid.

The kernel is constructed from the stress associated with the channel Stokeslet, including both pressure and viscous contributions. Its definition can be understood through an auxiliary problem: place a unit point force in direction \(i\) at the observation point \(\boldsymbol x\), and evaluate the resulting traction at \(\boldsymbol x'\), using \(\boldsymbol n(\boldsymbol x')\). The negative of the \(j\)th component of this traction defines \(K_{ij}(\boldsymbol x,\boldsymbol x')\). This is the negative reciprocal traction construction; its component expressions and their derivation are given in appendix~\ref{app:double-layer}.

Equation~\eqref{eq:double-layer-field} superposes the elementary double-layer fields over the cylinder boundary. The density \(\boldsymbol\phi\) controls their strengths and is determined as part of the solution. Although it has the dimensions of velocity, it is an auxiliary density: it is neither the prescribed cylinder velocity nor the physical surface traction. In particular, integrating \(\boldsymbol\phi\) does not give the hydrodynamic force.

\paragraph{Force and torque completion.} A double-layer field alone cannot represent the general flow driven by cylinder motion, because its resultant hydrodynamic force and torque on the cylinder vanish. We therefore supplement it with a channel Stokeslet and a channel rotlet placed at the cylinder centre:
\begin{equation}
\label{eq:completed-representation}
\boldsymbol u(\boldsymbol x)
=
\int_\Gamma
\boldsymbol K(\boldsymbol x,\boldsymbol x')
\,\boldsymbol\phi(\boldsymbol x')\,\mathrm ds'
+
\boldsymbol S(\boldsymbol x,\boldsymbol x_c)
\,\boldsymbol f^{c}
+
\boldsymbol r(\boldsymbol x,\boldsymbol x_c)
\,\tau^{c}.
\end{equation}
The matrix \(\boldsymbol S\) is the channel Stokeslet defined in \eqref{eq:channel-stokeslet}. The vector \(\boldsymbol r\) is the velocity generated by a positive unit point torque about the \(z\)-axis, with both plate boundary conditions incorporated. Its construction from source derivatives of \(\boldsymbol S\) is given in appendix~\ref{app:double-layer}.

The completion strengths \(\boldsymbol f^{c}\) and \(\tau^{c}\) are additional unknowns representing the strengths of the interior channel Stokeslet and rotlet. Momentum and angular-momentum balances show that they are the negatives of the hydrodynamic force and torque exerted by the fluid on the cylinder, and given by:
\begin{equation}
\label{eq:completion-loads}
\boldsymbol{f}^c=-\boldsymbol{F},
\qquad
\tau^c=-T.
\end{equation}
\noindent  The double-layer contribution carries no net force or torque. The derivation is given in appendix~\ref{app:double-layer}.

The completion singularities lie inside the solid cylinder, outside the physical fluid domain. Their role is to provide the missing resultant force and torque; the combined field remains a solution of the homogeneous Stokes equations throughout the fluid. All terms in \eqref{eq:completed-representation} satisfy the stationary-wall conditions.

\paragraph{Imposing the cylinder velocity.} To determine the unknown density and completion strengths, we take the limit of \eqref{eq:completed-representation} as the observation point approaches the cylinder from the fluid side. The double-layer potential has a boundary jump. With the kernel and normal conventions adopted here,
\begin{equation}
\label{eq:double-layer-jump}
\lim_{\epsilon\downarrow0}
\boldsymbol u^{D}
\bigl(\boldsymbol x_b+\epsilon\boldsymbol n(\boldsymbol x_b)\bigr)
=
-\frac12\boldsymbol\phi(\boldsymbol x_b)
+
\operatorname{p.v.}\!\int_\Gamma
\boldsymbol K(\boldsymbol x_b,\boldsymbol x')
\,\boldsymbol\phi(\boldsymbol x')\,\mathrm ds',
\qquad \boldsymbol x_b\in\Gamma.
\end{equation}
The principal value is defined by excluding a small boundary segment around \(\boldsymbol x_b\) and then letting its length tend to zero. The local term \(-\boldsymbol\phi/2\) accounts for the fluid-side limit and must be retained separately from the boundary integral.

The jump is determined by the local free-space singularity of the kernel. The wall correction is smooth across the cylinder boundary and therefore does not alter the jump coefficient. Appendix~\ref{app:double-layer} derives the sign in \eqref{eq:double-layer-jump} using the cylinder-to-fluid normal.

The completion fields are smooth on \(\Gamma\). Substituting the prescribed rigid-body velocity from \eqref{eq:cylinder-velocity} therefore gives
\begin{equation}
\label{eq:completed-boundary-equation}
\begin{aligned}
-\frac12\boldsymbol\phi(\boldsymbol x_b)
&+
\operatorname{p.v.}\!\int_\Gamma
\boldsymbol K(\boldsymbol x_b,\boldsymbol x')
\,\boldsymbol\phi(\boldsymbol x')\,\mathrm ds'
\\
&+
\boldsymbol S(\boldsymbol x_b,\boldsymbol x_c)
\,\boldsymbol f^{c}
+
\boldsymbol r(\boldsymbol x_b,\boldsymbol x_c)
\,\tau^{c}
\\
&=
U\boldsymbol e_x+V\boldsymbol e_y
+a\Omega\boldsymbol t(\boldsymbol x_b),
\qquad \boldsymbol x_b\in\Gamma.
\end{aligned}
\end{equation}
For prescribed \(U\), \(V\) and \(\Omega\), the unknowns are the boundary density \(\boldsymbol\phi\) and the three scalar completion strengths \(f^c_{x}\), \(f^c_y\) and \(\tau^c\). The boundary equation alone does not determine the density uniquely: adding any of the three density modes corresponding to planar rigid-body motion—two translations and one rotation—leaves the exterior velocity field unchanged. We remove this freedom by imposing
\begin{equation}
\label{eq:density-constraints}
\int_\Gamma\boldsymbol\phi\,\mathrm ds=\boldsymbol0,
\qquad
\int_\Gamma\boldsymbol t\boldsymbol{\cdot}\boldsymbol\phi\,\mathrm ds=0.
\end{equation}
The first relation supplies two scalar constraints that remove the translational modes. The second removes the rotational mode, whose boundary dependence is proportional to \(\boldsymbol t\), since \(\boldsymbol e_z\times(\boldsymbol x-\boldsymbol x_c)=a\boldsymbol t\). These constraints remove the remaining freedom in the auxiliary density without imposing zero physical force or torque; the resultant loads are carried by the completion strengths.

Equations~\eqref{eq:completed-boundary-equation} and \eqref{eq:density-constraints} are solved simultaneously. Discretising the boundary at \(N\) points gives \(2N\) density components and three completion strengths, matched by \(2N\) boundary-velocity equations and three scalar constraints. The resulting linear system is described in \S\ref{sec:discretisation}. Its solution determines the auxiliary density and completion strengths, from which the  velocity field follows through \eqref{eq:completed-representation}.

\subsection{Discretisation and resistance extraction}
\label{sec:discretisation}

We discretise the cylinder boundary using \(N\) points \(\boldsymbol x_j=\boldsymbol x_b(\theta_j)\), with associated arclength quadrature weights \(w_j\). Boundary integrals are approximated by weighted sums,
\begin{equation}
\label{eq:boundary-quadrature}
\int_\Gamma \boldsymbol g(\boldsymbol x')\,\mathrm ds'
\approx
\sum_{j=1}^{N}w_j\boldsymbol g(\boldsymbol x_j).
\end{equation}
Uniform angular spacing gives \(w_j=2\pi a/N\). For narrow gaps, a smooth mapping concentrates the boundary points near the closest wall, and the weights include the corresponding arclength Jacobian. The mapping and kernel-evaluation details are given in appendix~\ref{app:numerics}.

The unknown density is represented by its values at these points. We collect its components and the completion strengths into
\begin{equation}
\label{eq:discrete-unknowns}
\boldsymbol\Phi
=
\begin{pmatrix}
\phi_{x,1}\\
\phi_{y,1}\\
\vdots\\
\phi_{x,N}\\
\phi_{y,N}
\end{pmatrix},
\qquad
\boldsymbol\lambda
=
\begin{pmatrix}
f^c_x\\
f^c_y\\
\tau^c/a
\end{pmatrix}.
\end{equation}
Here, \(\boldsymbol\Phi\) contains the \(2N\) density components, and $\boldsymbol\lambda$ contains the three completion strengths. The density determines the double-layer contribution for both translational and rotational motion; no separate rotational density is required. The net torque is carried by the rotlet completion, whose unknown strength is represented by \(\tau^c/a\). Thus \(\boldsymbol\lambda\) contains the completion strengths, which are negatives of the hydrodynamic loads defined in \S\ref{sec:hydrodynamic-loads}.

Enforcing the boundary velocity equation at the same \(N\) points gives a Nyström discretisation of the integral equation. Together with the density constraints, it yields
\begin{equation}
\label{eq:discrete-system}
\begin{pmatrix}
\boldsymbol A & \boldsymbol B\\
\boldsymbol C & \boldsymbol 0
\end{pmatrix}
\begin{pmatrix}
\boldsymbol\Phi\\
\boldsymbol\lambda
\end{pmatrix}
=
\begin{pmatrix}
\boldsymbol b\\
\boldsymbol 0
\end{pmatrix},
\end{equation}
where \(\boldsymbol b\) contains the prescribed velocity components at the boundary points. The matrices \(\boldsymbol A\), \(\boldsymbol B\) and \(\boldsymbol C\) have dimensions \(2N\times2N\), \(2N\times3\) and \(3\times2N\), respectively. The complete system therefore contains \(2N+3\) equations and unknowns.

The matrix \(\boldsymbol A\) maps the density values to the double-layer boundary velocity. Its \(2\times2\) blocks are
\begin{equation}
\label{eq:density-block}
\boldsymbol A_{ij}
=
w_j\boldsymbol K_{ij}
-\frac12\delta_{ij}\boldsymbol I_2,
\end{equation}
where \(\delta_{ij}\) is the Kronecker delta and \(\boldsymbol I_2\) is the two-dimensional identity matrix. For distinct points, \(\boldsymbol K_{ij}=\boldsymbol K(\boldsymbol x_i,\boldsymbol x_j)\). At coincident points, the analytically determined boundary limit is used. For the circular boundary, the free-space contribution has a finite diagonal limit, to which the smooth wall correction is added. This kernel value and the explicit jump term \(-\boldsymbol I_2/2\) are distinct contributions.

The matrix \(\boldsymbol B\) gives the boundary velocity generated by the completion strengths. Its two-row block at observation point \(i\) is
\begin{equation}
\label{eq:completion-block}
\boldsymbol B_i
=
\left[
\boldsymbol S(\boldsymbol x_i,\boldsymbol x_c)\boldsymbol e_x
\quad
\boldsymbol S(\boldsymbol x_i,\boldsymbol x_c)\boldsymbol e_y
\quad
a\,\boldsymbol r(\boldsymbol x_i,\boldsymbol x_c)
\right].
\end{equation}
The factor \(a\) in the third column is required because the third unknown in \(\boldsymbol\lambda\) is \(\tau^c/a\). Hence this column contributes precisely \(\boldsymbol r \tau^c\) to the velocity.

Finally, the matrix \(\boldsymbol C\) imposes the three density constraints. Its \(3\times2\) block associated with source point \(j\) is
\begin{equation}
\label{eq:constraint-block}
\boldsymbol C_j
=
\frac{w_j}{2\pi a}
\begin{pmatrix}
1&0\\
0&1\\
-\sin\theta_j&\cos\theta_j
\end{pmatrix}.
\end{equation}
The first two rows impose zero boundary averages of the density components, while the third imposes zero average tangential density. Division by the circumference \(2\pi a\) normalises these equations without changing the constraints.

For a fixed geometry, the system matrix in \eqref{eq:discrete-system} is independent of the prescribed motion. The resistance coefficients are obtained by solving for three independent motions,
\begin{equation}
\label{eq:unit-motions}
(U,V,a\Omega)
=
(U_0,0,0),\qquad
(0,U_0,0),\qquad
(0,0,U_0),
\end{equation}
where \(U_0\) is an arbitrary reference speed. The corresponding boundary velocities at point \(j\), arranged as three columns, are
\begin{equation}
\label{eq:unit-boundary-velocities}
U_0
\begin{pmatrix}
1&0&-\sin\theta_j\\
0&1&\cos\theta_j
\end{pmatrix}.
\end{equation}
The third column therefore prescribes \(\Omega=U_0/a\). All three problems can be solved together using the same matrix factorisation.

Let \(\boldsymbol\lambda^{(U)}\), \(\boldsymbol\lambda^{(V)}\) and \(\boldsymbol\lambda^{(\Omega)}\) denote the resulting completion-strength vectors. The dimensionless hydrodynamic resistance matrix appearing in \eqref{eq:dimensionless-resistance} is
\begin{equation}
\label{eq:resistance-extraction}
\widehat{\boldsymbol R}
=
-\frac{1}{\mu U_0}
\left[
\boldsymbol\lambda^{(U)}
\quad
\boldsymbol\lambda^{(V)}
\quad
\boldsymbol\lambda^{(\Omega)}
\right].
\end{equation}
The minus sign converts the required completion strengths into the hydrodynamic loads used in \S\ref{sec:hydrodynamic-loads}. The resistance coefficients are therefore obtained directly from the completion unknowns, without additional surface-traction integration.

All matrix entries are calculated independently. Reciprocal symmetry and the entries that vanish by reflection symmetry are therefore available as checks on the numerical solution, rather than being imposed during assembly.

\subsection{Calculation of the hydraulic resistance for the pump problem}
\label{sec:pump-calculation}

The pump characteristic requires the additional hydraulic resistance \(C\), which is obtained by solving for a stationary cylinder in an imposed Poiseuille flow. This calculation uses the same completed boundary-integral operator as the rigid-motion problems, with a different prescribed boundary velocity.

For a nonzero flux \(Q\), write
\begin{equation}
\label{eq:pump-background-disturbance}
\boldsymbol u=\boldsymbol u_P+\boldsymbol u',
\quad
p=p_P+p',
\quad
\boldsymbol u_P(y)=\frac{6Q}{H^3}y(H-y)\boldsymbol e_x,
\quad
p_P(x)=-\frac{12\mu Q}{H^3}x.
\end{equation}
The disturbance \(\boldsymbol u'\) satisfies homogeneous Stokes equations, vanishes on the channel walls and decays as \(x\to\pm\infty\). Since the cylinder is held fixed, its boundary condition is
\begin{equation}
\label{eq:fixed-cylinder-disturbance}
\boldsymbol u'(\boldsymbol x_b)=-\boldsymbol u_P(y_b),
\qquad \boldsymbol x_b\in\Gamma.
\end{equation}
The calculation therefore remains posed in the infinite channel. The imposed pressure gradient is contained entirely in the known background field \(p_P\), while the disturbance pressure approaches finite constants at both ends.

Representing \(\boldsymbol u'\) by the completed double layer and applying the same density constraints gives
\begin{equation}
\label{eq:pump-discrete-system}
\begin{pmatrix}
\boldsymbol A & \boldsymbol B\\
\boldsymbol C & \boldsymbol 0
\end{pmatrix}
\begin{pmatrix}
\boldsymbol\Phi^{(Q)}\\
\boldsymbol\lambda^{(Q)}
\end{pmatrix}
=
\begin{pmatrix}
\boldsymbol b^{(Q)}\\
\boldsymbol 0
\end{pmatrix},
\qquad
\boldsymbol b_j^{(Q)}
=
-\frac{6Q}{H^3}y_j(H-y_j)
\begin{pmatrix}1\\0\end{pmatrix}.
\end{equation}
Here \(\boldsymbol\Phi^{(Q)}\) contains the two components of the disturbance density at each boundary node. The bold symbol \(\boldsymbol C\) denotes the constraint matrix defined in \eqref{eq:constraint-block} and is distinct from the scalar hydraulic-resistance coefficient \(C\). The completion vector is
\begin{equation}
\label{eq:pump-completion-vector}
\boldsymbol\lambda^{(Q)}
=
\begin{pmatrix}
f_{x}^{c,(Q)}\\
f_{y}^{c,(Q)}\\
\tau^{c,(Q)}/a
\end{pmatrix}.
\end{equation}
As in \eqref{eq:completion-loads}, the corresponding hydrodynamic loads are the negatives of these completion strengths. All matrix blocks are identical to those used in \S\ref{sec:discretisation}. Thus, the imposed-flux problem adds a fourth right-hand side and can reuse the same matrix factorisation. A convenient normalisation is \(Q=HU_0\), for which the mean background velocity is the reference speed \(U_0\).

The resulting density and completion strengths determine the excess pressure jump through the pressure functional derived in appendix~\ref{app:pressure}. Equivalently, the streamwise momentum equation gives
\begin{equation}
\label{eq:pump-pressure-extraction}
\Delta p_{\mathrm{ex}}(0,Q)
=p'_- -p'_+
=-\mu\,\partial_y^2\widehat{u'}(0,y),
\qquad
C=\frac{H^2}{\mu Q}\Delta p_{\mathrm{ex}}(0,Q),
\end{equation}
where \(\widehat{u'}\) is the Fourier transform of the streamwise disturbance velocity, evaluated on a horizontal line below or above the cylinder. In practice, the pressure functional is evaluated directly from the boundary solution, without reconstructing pressure at distant locations. Linearity makes \(C\) independent of the chosen nonzero \(Q\).

Combining this fixed-cylinder solution with the zero-flux rotational solution yields
\begin{equation}
\label{eq:computed-pump-characteristic}
\Delta p_{\mathrm{ex}}(\Omega,Q)
=
\frac{\mu a\Omega}{H}P^\Omega
+\frac{\mu Q}{H^2}C.
\end{equation}
The coefficient \(P^\Omega\) is supplied by the rotational problem, whereas \(C\) requires the additional solve above. For comparison with a finite-channel numerical solution, the empty-channel pressure drop \(12\mu QL/H^3\) is then added between measurement sections separated by \(L\), provided those sections lie sufficiently far from the cylinder for the flow to be developed.
\section{Finite-element calculations}
\label{sec:finite-element}

Independent finite-element calculations were performed using COMSOL Multiphysics 6.4 to assess the boundary-integral predictions. The two-dimensional incompressible Stokes equations were solved using the Creeping Flow formulation, with inertial terms omitted.

\subsection{Computational domain and boundary conditions}

The computational domain was a rectangular channel,
\begin{equation}
\label{eq:finite-domain}
-\frac{L}{2}\leq x\leq\frac{L}{2},
\qquad 0\leq y\leq H,
\end{equation}
excluding the circular cylinder of radius \(a\), centred at \((0,y_c)\). The zero-flux resistance calculations shown in figures~\ref{fig:resistance-a010}--\ref{fig:resistance-a045} covered \(a/H=0.1\), \(0.3\) and \(0.45\), with \(L/H=20\). Additional calculations at \(L/H=100\) were used to assess sensitivity to the streamwise extent. The pressure-response and pumping comparisons also included \(a/H=0.2\) and \(0.4\); the through-flow finite-element calculations used \(L/H=10\). Throughout, \(H=1\) and \(\mu=1\) were used, and the loads and pressure differences were normalised according to \S\ref{sec:hydrodynamic-loads} and \S\ref{sec:viscous-pump}.

For the three rigid-body resistance problems, no-slip conditions were imposed on the upper and lower plates and on both streamwise end boundaries. The closed ends enforce zero net flux at every cross-section not intersecting a translating cylinder, approximating the zero-through-flow far-field condition of the infinite-channel resistance problem. A separate set of through-flow calculations, described below, used streamwise end conditions admitting a prescribed flux \(Q\) instead of closed ends. The prescribed rigid-body velocity was applied on the cylinder:
\begin{equation}
\label{eq:fem-cylinder-velocity}
u=U-\Omega(y-y_c),
\qquad
v=V+\Omega x.
\end{equation}
Each calculation represents the instantaneous Stokes flow at the specified cylinder position. A pressure-point constraint at the uppermost point of the cylinder fixed the arbitrary pressure reference.

For each cylinder position, three separate problems prescribed unit parallel translation, unit normal translation and unit angular velocity, with the other two components set to zero. Positions were selected in the lower half of the channel, \(a<y_c\leq H/2\), including near-wall clearances and the centreline. Only lower-half configurations are necessary because reflection about \(y=H/2\) determines the corresponding upper-half coefficients.

\subsection{Discretisation and solution}

Quadratic interpolation was used for both velocity and pressure, corresponding to equal-order \(P_2+P_2\) elements in the COMSOL Creeping Flow interface. Streamline diffusion was enabled for pressure stabilisation. The domain was discretised using a free triangular mesh, supplemented by boundary-layer meshing with two layers along the cylinder and selected channel boundaries. The stationary equations were solved using the PARDISO direct solver.

The cylinder boundary was refined according to the minimum wall clearance \(h=y_c-a\) for the lower-half positions considered. Its prescribed maximum element size was
\begin{equation}
\label{eq:fem-cylinder-mesh}
\ell_{\Gamma,\max}
=\min\left(\frac{a}{20},\frac{h}{3}\right).
\end{equation}
This rule increases the boundary resolution as the gap decreases. For the \(a/H=0.1\) refinement comparison, the central region \(|x|\leq H\) used a maximum element size of \(H/20\), while the outer channel regions used \(H/10\). A maximum element growth rate of 1.15 provided a gradual transition between the refined and outer regions. This gap-dependent boundary refinement, rather than a fixed mesh size, was used to resolve the increasingly narrow clearances.

\subsection{Load extraction and numerical checks}

The force components and torque per unit cylinder length were obtained by integrating the total surface traction around the cylinder, with the torque evaluated about \((0,y_c)\). The reaction loads calculated in COMSOL were converted to the hydrodynamic-load convention and normalisation defined in \S\ref{sec:hydrodynamic-loads}, in which the force and torque are those exerted by the fluid on the cylinder. The force generated by rotation and the torque generated by parallel translation were evaluated independently, allowing reciprocal symmetry to be checked without imposing it during post-processing.

The pressure difference was obtained from averages over the two channel ends:
\begin{equation}
\label{eq:fem-pressure-difference}
\Delta p_L
=\overline p_- -\overline p_+,
\qquad
\overline p_\pm
=\frac{1}{H}\int_0^H p(\pm L/2,y)\,\mathrm dy.
\end{equation}
This difference is independent of the pressure reference. For sufficiently distant end boundaries, it approaches the pressure jump of the infinite-channel solution. The pressure-response coefficients were extracted independently from the unit parallel-translation and unit-rotation solutions as
\begin{equation}
\label{eq:fem-pressure-coefficients}
P^U=\frac{H\,\Delta p_L^{(U)}}{\mu U},
\qquad
P^\Omega=\frac{H\,\Delta p_L^{(\Omega)}}{\mu a\Omega},
\end{equation}
with the inactive motion components set to zero in each case. Normal translation provided the additional symmetry check \(P^V=0\).

For the excess hydraulic resistance, the cylinder was held stationary and a separate through-flow problem was solved with prescribed volume flux per unit span \(Q\). The cylinder and the plates remained no-slip boundaries, while the streamwise ends admitted through-flow chosen to reproduce fully developed Poiseuille flow away from the cylinder. Specifically, the inlet used a fully developed Poiseuille profile with flow rate \(Q=1\), and the outlet used a normal-flow condition with static pressure \(p_0=0\). Unlike the closed-end calculations, the measured pressure difference contains both the empty-channel pressure drop and the contribution of the obstacle. Accordingly,
\begin{equation}
\label{eq:fem-excess-resistance}
C_{\mathrm{FEM}}
=\frac{H^2}{\mu Q}
\left(\Delta p_L^{(Q)}-\frac{12\mu QL}{H^3}\right),
\qquad
\Delta p_L^{(Q)}=\overline p_-^{(Q)}-\overline p_+^{(Q)},
\end{equation}
where \(\Omega=U=V=0\) for this calculation and the pressure averages are taken at the two end sections. Thus \(C_{\mathrm{FEM}}\) measures the \emph{excess} resistance of the cylinder, rather than the total resistance of the finite channel. In figures~\ref{fig:pump-performance}(d)--(f), the finite-element pump characteristics are constructed from the independently computed resistance and pressure coefficients using the relations of \S\ref{sec:viscous-pump}; they are not separate finite-element simulations at each pump operating point.

Numerical sensitivity was assessed through comparisons of channel lengths and local mesh configurations. In the recorded length comparison at \(L/H=20\) and \(100\), the nonzero resistance coefficients agreed with the corresponding boundary-integral values to within \(0.0081\%\). At \(L/H=20\), introducing the refined central region changed these coefficients by less than \(0.016\%\). The centreline translation–rotation coupling was excluded from relative comparisons because its theoretical value is zero. These estimates reflect the precision of the exported numerical data.

Additional checks used reciprocal symmetry and the coefficients that vanish by reflection symmetry. The latter were assessed through their absolute magnitudes rather than relative errors. Comparisons with the boundary-integral solution and the leading lubrication predictions are presented in §\ref{sec:resistance-results}.

\section{Results}
\subsection{Resistance coefficients and translation--rotation coupling}
\label{sec:resistance-results}

\begin{figure}[!htbp]
\centering
\includegraphics[width=\textwidth]{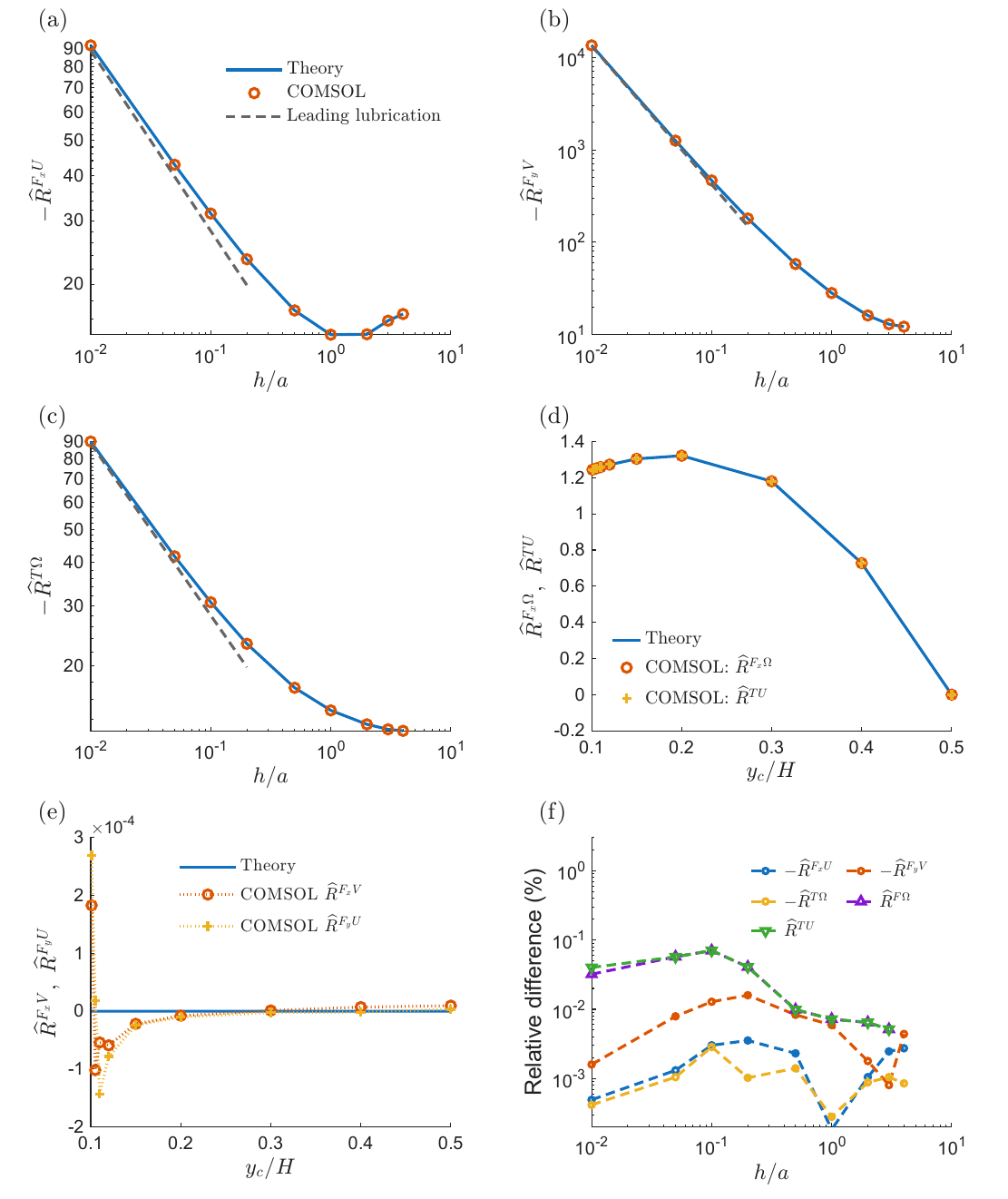}
\captionsetup{skip=2pt}
\caption{Resistance coefficients for \(a/H=0.1\). (a) Parallel translation; (b) normal translation; (c) rotation; (d) reciprocal translation--rotation coupling; (e) symmetry-forbidden translation couplings; (f) relative differences between the boundary-integral and finite-element results for the nonzero coefficients. The diagonal coefficients are shown with their signs reversed. Solid curves denote completed double-layer results, symbols denote COMSOL calculations, and dashed curves in panels (a)--(c) show the leading single-wall lubrication estimates.}
\label{fig:resistance-a010}
\end{figure}

\begin{figure}[!htbp]
\centering
\includegraphics[width=\textwidth]{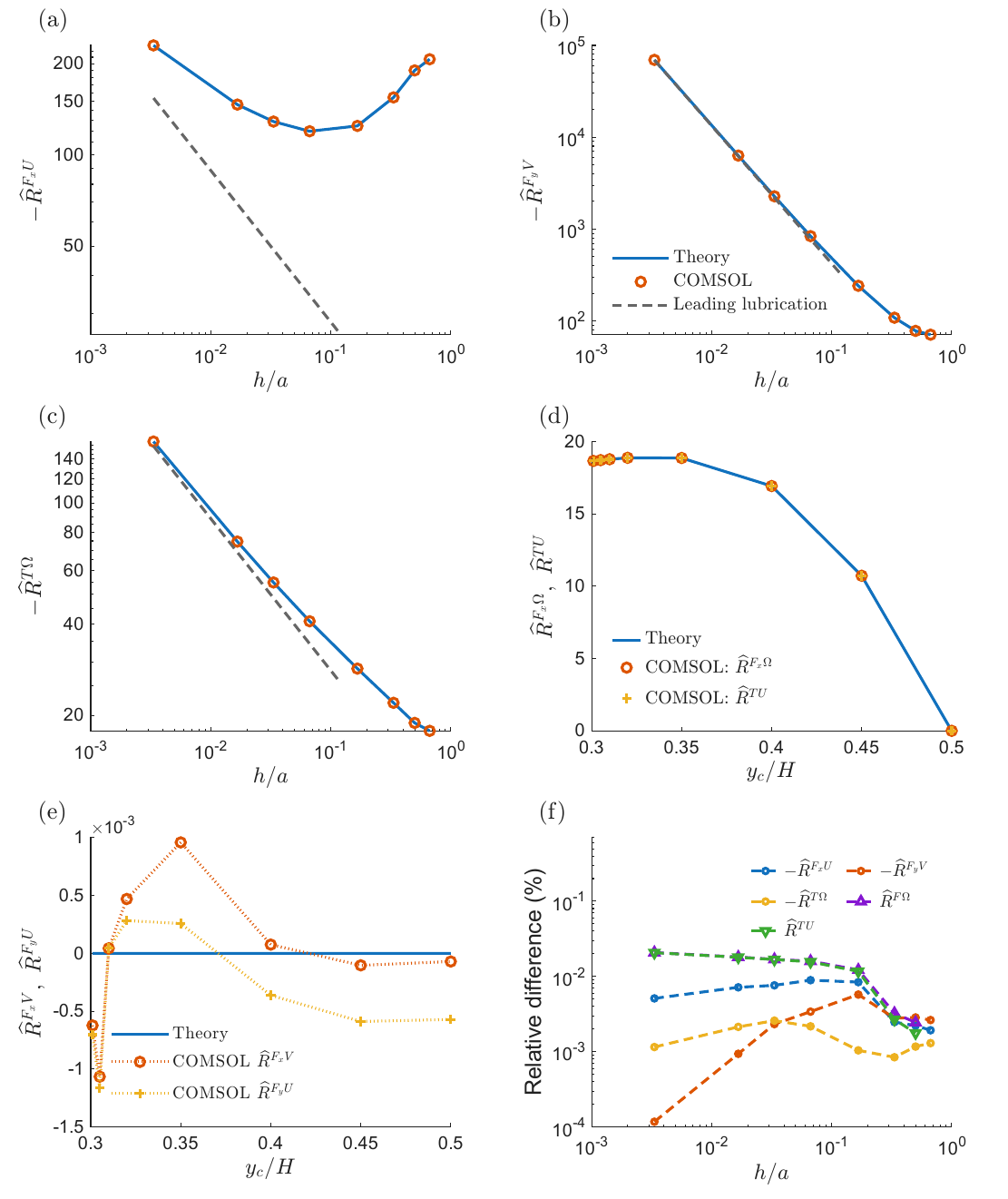}
\captionsetup{skip=2pt}
\caption{Resistance coefficients for \(a/H=0.3\). (a) Parallel translation; (b) normal translation; (c) rotation; (d) reciprocal translation--rotation coupling; (e) symmetry-forbidden translation couplings; (f) relative differences between the boundary-integral and finite-element results for the nonzero coefficients. The diagonal coefficients are shown with their signs reversed. Curves, symbols, and dashed lubrication estimates are as described in figure~\ref{fig:resistance-a010}.}
\label{fig:resistance-a030}
\end{figure}

\begin{figure}[!htbp]
\centering
\includegraphics[width=\textwidth]{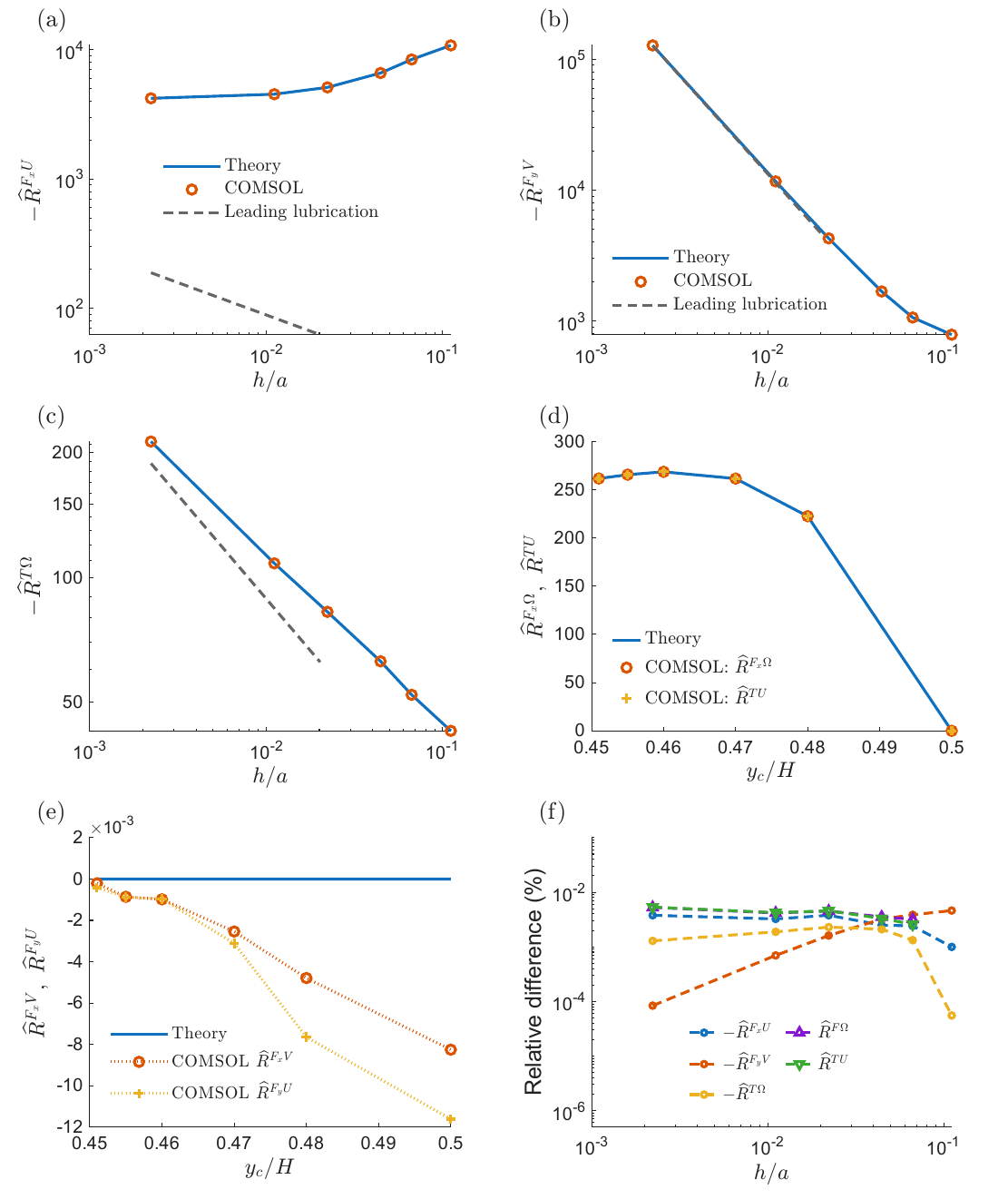}
\captionsetup{skip=2pt}
\caption{Resistance coefficients for \(a/H=0.45\). (a) Parallel translation; (b) normal translation; (c) rotation; (d) reciprocal translation--rotation coupling; (e) symmetry-forbidden translation couplings; (f) relative differences between the boundary-integral and finite-element results for the nonzero coefficients. The diagonal coefficients are shown with their signs reversed. Curves, symbols, and dashed lubrication estimates are as described in figure~\ref{fig:resistance-a010}.}
\label{fig:resistance-a045}
\end{figure}

Figures~\ref{fig:resistance-a010}--\ref{fig:resistance-a045} compare the dimensionless resistance coefficients obtained from the completed double-layer formulation with independent finite-element calculations for \(a/H=0.1\), \(0.3\) and \(0.45\). The comparisons cover parallel translation, normal translation and rotation, with cylinder positions ranging from narrow lower-wall clearances to the channel centreline. They assess both the nonzero coefficients and the cancellation of symmetry-forbidden forces, while the independently calculated translation--rotation coefficients provide a reciprocal consistency check. The results also distinguish the local resistance associated with the nearest wall from the response imposed by confinement and the zero-flux condition. This distinction is evident in the different positional dependences of the drag, torque and coupling coefficients, which determine the force-free translation speed and the torque required to sustain rotation.

The boundary-integral and finite-element results are nearly indistinguishable on the coefficient plots. The relative differences shown for the nonzero coefficients remain below \(0.1\%\) across all three confinement ratios; for \(a/H=0.45\), they are below \(0.01\%\). Agreement is therefore maintained as the cylinder diameter increases from one fifth to nine tenths of the channel height and the resistance changes by several orders of magnitude. These differences measure agreement between the two numerical methods rather than an independently established error bound. At the centreline, where the translation--rotation coupling vanishes, its accuracy must instead be assessed through an absolute residual, since a relative difference with respect to zero is undefined.

The vanishing transverse force during parallel translation supplies a particularly stringent cancellation test. Reflection about \(x=0\), combined with Stokes linearity, requires \(R^{F_yU}=0\) at every cylinder height, including strongly eccentric positions. The reciprocal coefficient \(R^{F_x V}\) must also vanish. In the present method these coefficients are computed rather than imposed, and, relative to the corresponding diagonal coefficients, they remain below \(10^{-11}\) and \(3\times10^{-10}\), respectively, whereas the finite-element calculations produce small nonzero residuals. When normalised by \(\mu\), their absolute magnitudes reach orders \(10^{-4}\), \(10^{-3}\) and \(10^{-2}\) for \(a/H=0.1\), \(0.3\) and \(0.45\), respectively. Although small compared with the nonzero resistances, these residuals are numerical symmetry breaking errors and do not represent physical lift. They mark a difference between the numerical results for a finite computational domain and the infinite channel represented exactly by the Green function.

The independently calculated coefficients \(\widehat R^{F_x \Omega}\) and \(\widehat R^{TU}\) agree closely for all three confinement ratios. They describe distinct prescribed motions---the parallel force on a rotating cylinder and the torque on a translating cylinder---so their agreement tests the reciprocal structure of the complete force--torque response. Both coefficients vanish at the centreline, as required by reflection symmetry about \(y=H/2\). Below the centreline, the coupling is positive and exhibits a shallow off-centre maximum before decreasing to zero. The sampled maxima occur near \(y_c/H=0.20\), \(0.35\) and \(0.46\), respectively, while their magnitudes increase from order unity to order \(10\) and order \(10^2\). Confinement therefore strengthens the dimensionless coupling substantially, even though symmetry always eliminates it at the centreline.

The leading single-wall lubrication estimates provide a different test, concerned with the physical origin of the resistance. Normal translation produces the strongest local singularity, \(-\widehat R^{F_yV}\propto(h/a)^{-3/2}\), and its near-wall resistance follows the leading estimate closely for all three cylinder sizes. Parallel translation and rotation have weaker single-wall singularities, proportional to \((h/a)^{-1/2}\), but their departures from these estimates differ markedly. The rotational resistance approaches its leading near-wall behaviour while retaining finite-clearance corrections. The parallel resistance is much more strongly affected by confinement: for \(a/H=0.45\), it remains more than an order of magnitude above the leading single-wall estimate even at the smallest plotted clearance. A small value of \(h/a\) is thus insufficient to guarantee that the leading single-wall expression approximates every component of the channel resistance. The opposite clearance, \((H-2a-h)/a\), and the imposed mode of motion also matter. Since the dashed curves contain only the leading single-wall term, departures at finite clearance include higher-order local corrections as well as the influence of the second plate.

The positional dependence of the parallel drag makes the global contribution especially clear. For \(a/H=0.1\) and \(0.3\), \(-\widehat R^{F_x U}\) first decreases as the cylinder moves away from the lower wall, reaches an off-centre minimum, and then increases towards the centreline. For \(a/H=0.45\), it increases towards the centreline over the sampled positions; the data do not resolve a separate interior minimum. In contrast, the normal and rotational resistances decrease towards their centreline values. The non-monotonic parallel drag reflects competition between the local near-wall stresses and the resistance to the return flow required by zero net flux. An eccentric position leaves a relatively open passage on one side, whereas centring a strongly confined cylinder divides the available clearance between two narrow passages. The resulting increase in the resistance to return flow can outweigh the reduction in the nearest-wall contribution. At fixed \(a/H<1/2\), however, the local parallel-translation singularity must ultimately dominate as \(h/a\to0\); the strongly confined data show that the plotted gaps need not yet lie in that asymptotic regime.

\begin{figure}[!htbp]
\centering
\includegraphics[width=\textwidth]{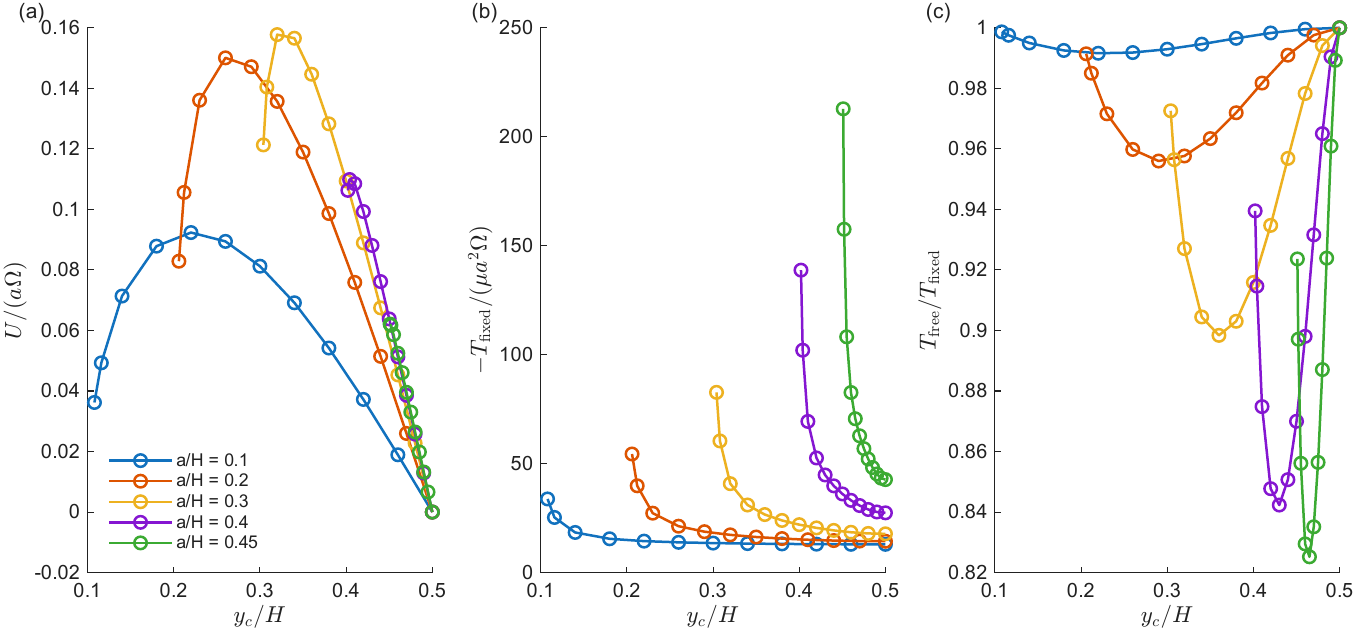}
\caption{Force-free translation of a rotating cylinder for \(a/H=0.1\), \(0.2\), \(0.3\), \(0.4\) and \(0.45\). (a) Induced parallel velocity \(U/(a\Omega)\). (b) Hydrodynamic torque \(-T_{fixed}/(\mu a^2\Omega)\) required to sustain rotation for a fixed cylinder. (c) Ratio of the torques for a free translating cylinder to a fixed cylinder. Curves show boundary-integral results, symbols show COMSOL results; colours identify the confinement ratios in the left-panel legend.}
\label{fig:force-free-translation}
\end{figure}

Figure~\ref{fig:force-free-translation} shows the force-free response for five confinement ratios, \(a/H=0.1\), \(0.2\), \(0.3\), \(0.4\) and \(0.45\). As established in \S\ref{sec:translation-rotation-coupling}, \(F_x=F_y=0\) gives \(V=0\) and \(U/(a\Omega)=-\widehat R^{F_x\Omega}/\widehat R^{F_xU}\). The positive coupling and negative parallel resistance therefore produce \(U/(a\Omega)>0\) for a cylinder rotating near the bottom wall. For positive \(\Omega\), no-slip rolling along the lower wall would require \(U/(a\Omega)=-1\); the computed positive ratio means that the force-free cylinder translates in the opposite direction. The induced translation vanishes at the centreline, where the coupling is zero.  Among the sampled positions, the maximum \(U/(a\Omega)\) increases from approximately \(0.09\) at \(a/H=0.1\) to \(0.15\)--\(0.16\) at \(a/H=0.2\)--\(0.3\), then decreases to approximately \(0.11\) and \(0.06\) at \(a/H=0.4\) and \(0.45\), respectively. The maximum force-free translation speed does not coincide with the maximum translation–rotation coupling because the speed depends on the ratio of the coupling coefficient to the parallel drag coefficient, both of which vary with cylinder position. Thus, stronger translation--rotation coupling does not imply a larger dimensionless translation speed: under strong confinement, its growth is outweighed by the increase in parallel drag.

Figure~\ref{fig:force-free-translation}(b) shows the normalised magnitude of the hydrodynamic torque on a cylinder whose centre is held fixed, $-T_{\mathrm{fixed}}/(\mu a^2\Omega)$. This quantity increases strongly near the lower wall and retains a substantial confinement dependence at the centreline. Allowing the cylinder to translate freely reduces the torque magnitude, as quantified by the ratio in figure~\ref{fig:force-free-translation}(c):

For the same geometry and \(\Omega\), comparison with a cylinder whose centre is held fixed gives
\begin{equation}
\label{eq:free-fixed-torque-ratio}
\frac{T_{\mathrm{free}}}{T_{\mathrm{fixed}}}
=1-\frac{\bigl(\widehat R^{F_x\Omega}\bigr)^2}
{\widehat R^{F_xU}\widehat R^{T\Omega}},
\qquad
T_{\mathrm{fixed}}=\mu a^2\widehat R^{T\Omega}\Omega.
\end{equation}
Negative definiteness of the resistance matrix ensures that this ratio is positive and no greater than unity. Therefore, allowing the cylinder to translate reduces the torque required to maintain a given angular velocity. The reduction is small for \(a/H=0.1\), but reaches approximately \(10\%\) for \(a/H=0.3\) and \(16\)--\(17\%\) for \(a/H=0.4\)--\(0.45\). The ratio returns to unity at the centreline because translation and rotation decouple there. At fixed angular velocity, the same ratio describes the reduction in mechanical power input, \(-T\Omega\). The contrast between the intermediate-confinement maximum in dimensionless speed and the larger torque reduction under strong confinement reflects the different combinations of resistance coefficients governing these two responses.

\subsection{Pressure generation and pumping performance}
\label{sec:pump-results}

\begin{figure}[!htbp]
\centering
\includegraphics[width=0.84\textwidth]{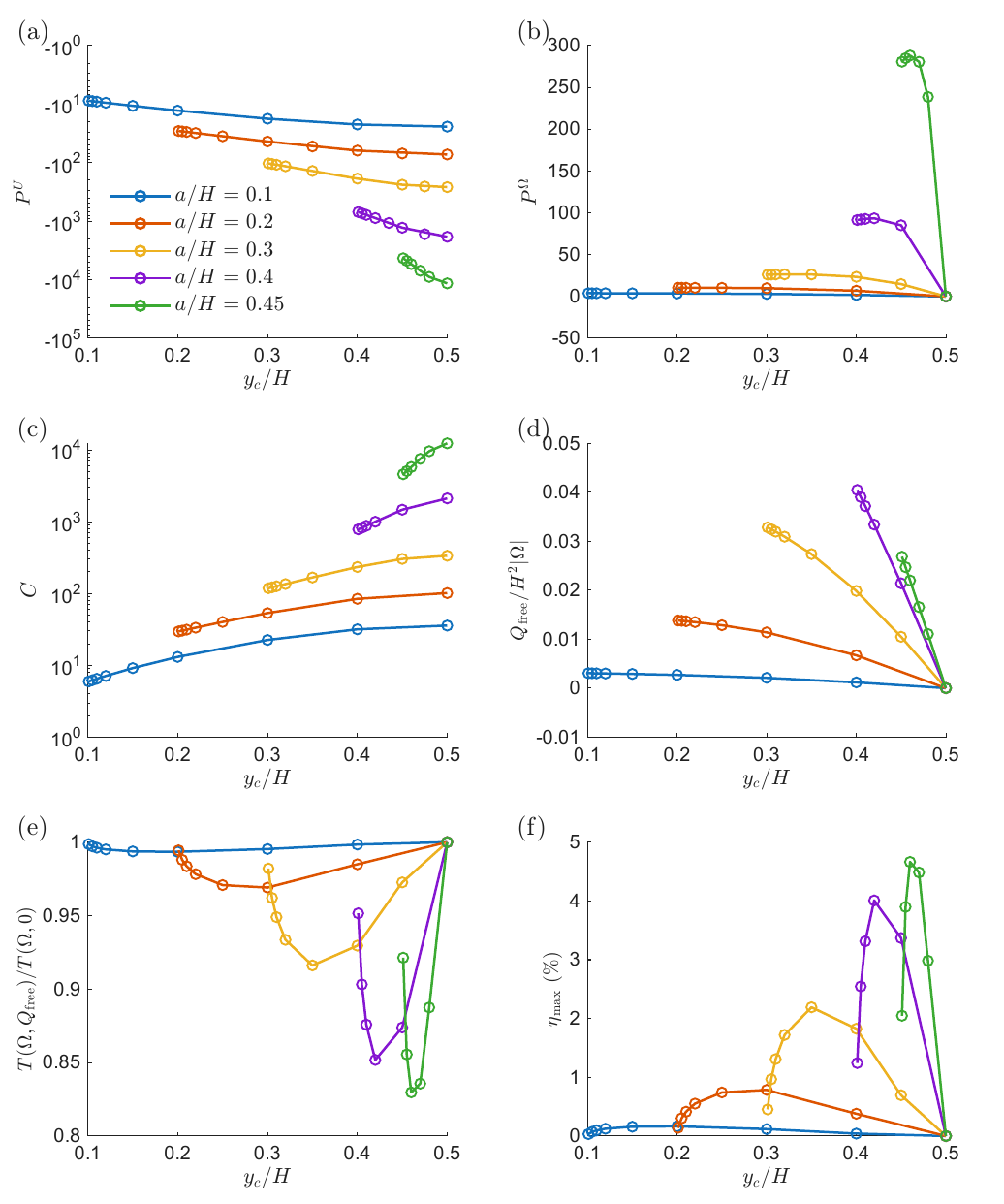}
\caption{Pressure response and pumping performance for \(a/H=0.1\), \(0.2\), \(0.3\), \(0.4\) and \(0.45\). (a) Zero-flow pressure coefficient for parallel translation, \(P^U\); (b) zero-flow rotational pressure coefficient, \(P^\Omega\); (c) excess hydraulic resistance, \(C\); (d) normalised free-delivery flow rate, \(Q_{\mathrm{free}}/(H^2|\Omega|)\); (e) driving torque at free delivery divided by its zero-flow value at the same angular velocity; and (f) maximum hydraulic efficiency, expressed as a percentage. Panels (d)--(f) use \(L/H=10\) and \(\Omega=-2\pi\) in the units of the computations, with the cylinder centre fixed. Solid lines denote theoretical predictions and symbols denote finite-element results; quantities in (d)--(f) are derived from the corresponding pressure and torque coefficients. Panel (a) uses logarithmic spacing in \(\lvert P^U\rvert\), with tick labels retaining the negative sign of \(P^U\); panel (c) uses a logarithmic scale.}
\label{fig:pump-performance}
\end{figure}

Figure~\ref{fig:pump-performance} shows the pressure response and pumping performance for \(a/H=0.1\)--\(0.45\). Panels (a)--(c) give the pressure coefficients for parallel translation and rotation at zero imposed flow, \(P^U\) and \(P^\Omega\), and the excess hydraulic resistance \(C\), respectively. The agreement between the boundary-integral predictions and the finite-element results is excellent. The maximum pointwise relative discrepancies are $0.0111\%$, $0.0118\%$ and $0.0122\%$ for $P^U$, $P^\Omega$ and $C$, respectively. Thus, the discrepancies remain below $0.0122\%$ for all three coefficients over the configurations considered, excluding points where the theoretical coefficients vanish by symmetry. 

Panels (d)--(f) describe pumping by a cylinder rotating about a fixed axis, with \(U=V=0\), at a common angular velocity and with pressure sections separated by \(L/H=10\). For the clockwise rotation considered here, a cylinder below the channel centreline pumps fluid in the positive \(x\)-direction. The theoretical curves and the results obtained from the finite-element coefficients agree closely at the scale of the figure.

 For each radius, \(\lvert P^U\rvert\) increases as the cylinder approaches the centreline, and its magnitude grows substantially with confinement. Parallel translation displaces fluid along the channel; under the zero-flow condition, compensating return flow must pass through the clearances around the cylinder. Moving the cylinder towards the centreline reduces the wider of these passages and increases the required pressure difference. Unlike the rotational response, \(P^U\) remains nonzero at the centreline. The coefficient for transverse translation, \(P^V\), vanishes by symmetry and is not shown.

The rotational pressure coefficient has a different positional dependence (figure~\ref{fig:pump-performance}(b)). It is positive below the centreline and vanishes at \(y_c/H=0.5\), where symmetry eliminates the axial pumping response when there is no net through-flow in the channel. For \(a/H=0.1\), $P^\Omega$ is largest at the smallest wall clearance. For the larger cylinders, a shallow maximum develops away from the wall before the coefficient decreases to zero at the centreline. Its largest sampled value increases from approximately \(3.85\) at \(a/H=0.1\) to \(288\) at \(a/H=0.45\). For the fixed clockwise rotation considered here, the delivered  shut-off pressure difference  \(-\Delta p_L(\Omega,0)\) is proportional to \(P^\Omega\) and  need not be plotted separately.

The excess hydraulic resistance \(C\) increases towards the centreline for each radius (figure~\ref{fig:pump-performance}(c)). An eccentric cylinder leaves a relatively open passage on one side, whereas a centred cylinder divides the available clearance between two equal gaps. The resistance to through-flow is therefore greatest at the centreline, where \(C\) increases from approximately \(36\) to \(1.23\times10^4\) over the range of radii considered. The relative contributions of the cylinder and the surrounding channel also change considerably. For \(a/H=0.1\), the empty-channel contribution \(12L/H=120\) exceeds \(C\) throughout the sampled range; for \(a/H=0.45\), the cylinder contribution dominates. Both contributions must be retained when relating the pressure generated by rotation to the delivered flow rate.

Figure~\ref{fig:pump-performance}(d) shows the resulting normalized free-delivery flow rate,\allowbreak $Q_{\mathrm{free}}/(H^2|\Omega|)$. Its dependence on \(P^\Omega/(C+12L/H)\), given by \eqref{eq:pump-free-delivery}, explains why the position of maximum pressure need not give the largest flow rate. The largest flow rate occurs at the smallest clearance for each radius, and the flow rate decreases towards zero at the centreline among the sampled results. Increasing the radius initially increases delivery, but this trend reverses between \(a/H=0.4\) and \(0.45\): the stronger rotational pressure response is outweighed by the increase in hydraulic resistance. Thus the largest free-delivery flow rate among the geometries considered occurs for \(a/H=0.4\), although the largest cylinder produces the highest shut-off pressure.

Figure~\ref{fig:pump-performance}(e) compares the driving torque at free delivery with its zero-flow value, \(T(\Omega,0)\), at the same angular velocity and with the cylinder centre fixed in both cases, for which the normalized values are shown in Figure~\ref{fig:force-free-translation}(b). Combining \eqref{eq:pump-free-delivery} and \eqref{eq:pump-torque} gives
\begin{equation}
\label{eq:pump-torque-ratio}
\frac{T(\Omega,Q_{\mathrm{free}})}
{T(\Omega,0)}
=1-\chi,
\qquad
\chi=\frac{(P^\Omega)^2}
{(-\widehat R^{T\Omega})(C+12L/H)}.
\end{equation}
The two torques have the same sign, so the ratio also compares their magnitudes. The induced through-flow reduces the torque required to sustain rotation. This reduction is small for \(a/H=0.1\), but reaches approximately \(17\%\) for \(a/H=0.45\). For each radius, the ratio has an off-centre minimum and returns to unity at the centreline, where \(P^\Omega=Q_{\mathrm{free}}=0\). Thus the greatest relative reduction in driving torque does not occur at the position of largest free-delivery flow rate.

The efficiency accounts for both the useful hydraulic power and the rotational power input,
\begin{equation}
\label{eq:pump-efficiency}
\eta(Q)=\frac{Q\,\Delta p_L(\Omega,Q)}
{T(\Omega,Q)\,\Omega},
\qquad 0\leq Q\leq Q_{\mathrm{free}}.
\end{equation}
The useful hydraulic power vanishes at both shut-off and free delivery, so the maximum efficiency occurs at an intermediate flow rate. Away from the centreline, setting \(s=Q/Q_{\mathrm{free}}\) in the pressure and torque relations gives
\begin{equation}
\label{eq:pump-efficiency-coupling}
\eta(s)=\frac{\chi s(1-s)}{1-\chi s},
\qquad
\eta_{\max}=\frac{1-\sqrt{1-\chi}}{1+\sqrt{1-\chi}}.
\end{equation}
Consequently, the positions of minimum torque ratio in panel (e) coincide with those of maximum \(\eta_{\max}\) in panel (f). Both quantities measure the strength of the rotation--through-flow coupling relative to the rotational and hydraulic resistances.

In figure~\ref{fig:pump-performance}(f), \(\eta_{\max}\) is obtained by varying the flow rate at each fixed cylinder position and angular velocity. Each radius has an off-centre position of highest efficiency. Close to the wall, the large driving torque limits efficiency; towards the centreline, the loss of pressure generation reduces the useful output. The largest sampled efficiencies are approximately \(0.16\%\), \(0.78\%\), \(2.19\%\), \(4.01\%\) and \(4.66\%\) for \(a/H=0.1\), \(0.2\), \(0.3\), \(0.4\) and \(0.45\), respectively. For the largest cylinder, the efficiency peaks near \(y_c/H=0.46\), corresponding to \(h/a\simeq0.022\). This position differs from the smallest sampled clearance, which gives the largest free-delivery flow rate for that radius. Among the radii examined, \(a/H=0.45\) yields the highest maximum efficiency, whereas \(a/H=0.4\) yields the largest free-delivery flow rate. These efficiencies include the resistance of the surrounding channel and refer to the specified \(L/H=10\).

The flow rate for the maximum efficiency is close to, but greater than, \(Q_{\mathrm{free}}/2\). The latter maximises hydraulic output power because the pressure--flow rate relation is linear. The input power, however, decreases with increasing flow rate at fixed angular velocity, shifting the efficiency maximum to a larger flow rate. Differentiating \eqref{eq:pump-efficiency}, using \eqref{eq:pump-total-pressure} and \eqref{eq:pump-torque}, gives
\begin{equation}
\label{eq:pump-efficiency-optimum}
\frac{Q_{\eta_{\max}}}{Q_{\mathrm{free}}}
=\frac{1}{1+\sqrt{1-\chi}}
=\frac{1+\eta_{\max}}{2},
\end{equation}
where \(\eta_{\max}\) is expressed as a fraction. At the sampled positions of highest efficiency for the five radii, \(Q_{\eta_{\max}}/Q_{\mathrm{free}}\) is approximately \(0.5008\), \(0.5039\), \(0.5110\), \(0.5200\) and \(0.5233\), respectively. Thus the optimum lies slightly above the half-flow value varying between \(0.16\%\) and \(4.66\%\) among the values sampled here. Even for \(a/H=0.45\), operating at \(Q_{\mathrm{free}}/2\) gives an efficiency of approximately \(4.654\%\), compared with \(4.664\%\) at the optimum. Half-flow therefore provides a close approximation to the most efficient operating condition over the parameter range considered, although the two optima are distinct.

\section{Conclusion}
A completed double-layer boundary-integral formulation has been developed for the resistance and pressure response of a circular cylinder between parallel plates. The channel Green function satisfies no slip on both plates, so only the cylinder boundary requires discretisation. Stokeslet and rotlet completions determine the resultant force and torque, while the pressure response is obtained from the zero-wavenumber contribution. The same formulation describes prescribed translation and rotation, force-free motion and viscous pumping, with imposed through-flow incorporated through an additional Poiseuille-flow problem. Excellent agreement with independent finite-element calculations is obtained for the resistance coefficients, the pressure generated by translation and rotation, and the excess hydraulic resistance over the confinement ratios and cylinder positions examined. The recovery of reciprocal coupling and symmetry-required cancellations provides further checks on the formulation.

The results demonstrate the importance of resolving local gap stresses together with the return flow imposed by confinement. Although the leading single-wall approximation describes the near-wall resistance to normal translation well, its accuracy for parallel translation and rotation depends strongly on both confinement and cylinder position. In particular, the parallel resistance can increase towards the centreline despite the increasing distance from the nearest wall. This behaviour reflects the resistance to returning the displaced fluid through the two passages and cannot be inferred from the minimum clearance alone.

Under the zero-flow-rate condition, a rotating cylinder undergoes backward sliding in every eccentric configuration examined: its force-free translation is opposite to the direction of rolling along the nearer wall. Below the centreline, this corresponds to \(U/(a\Omega)>0\); the translation vanishes at the centreline by symmetry. The dimensionless speed has an off-centre maximum and varies non-monotonically with confinement. Its largest sampled values, approximately \(0.15\)–\(0.16\), occur at \(a/H=0.2\)–\(0.3\). Stronger confinement increases translation–rotation coupling but produces an even larger increase in parallel drag, reducing \(U/(a\Omega)\). Allowing translation also reduces the torque required to maintain rotation, with reductions reaching approximately \(16\)–\(17\%\) in the more strongly confined cases.

For a cylinder rotating about a fixed axis, pumping performance is governed by the combined pressure generation, hydraulic resistance and driving torque. Increasing confinement strengthens pressure generation but does not necessarily increase the free-delivery flow rate. At the prescribed angular velocity and \(L/H=10\), the largest sampled \(Q_{\mathrm{free}}\) occurs for \(a/H=0.4\), whereas \(a/H=0.45\) gives the highest maximum efficiency, approximately \(4.7\%\). The most efficient position lies away from both the wall and the centreline, balancing the near-wall torque requirement against the loss of pressure generation as the cylinder becomes centred. Maximum efficiency occurs at a flow rate slightly above \(Q_{\mathrm{free}}/2\), reaching approximately \(0.523Q_{\mathrm{free}}\) in the most efficient case. These findings distinguish the conditions that favour translation speed, pumping rate and efficiency, and demonstrate how the complete resistance and pressure response determines performance beyond the limits of single-wall or narrow-gap approximations.

\section*{Declaration of interests}
The authors report no conflict of interest.

\section*{Acknowledgements}
The authors acknowledge the use of OpenAI's ChatGPT (\url{https://chatgpt.com}), including GPT-5.6 Sol, GPT-6 Astra and GPT-6.1 Astra, as an interactive AI-assisted research tool during the development of this work (July--October 2026). ChatGPT was used to assist in exploring and developing the analytical formulation, including the Fourier-based channel Green function and completed double-layer boundary-integral method; developing, debugging and refining the MATLAB numerical implementation of the boundary-integral formulation; examining its convergence, reciprocal identities and comparisons with independently obtained finite-element results; and analysing the resistance, translation--rotation coupling and pumping characteristics. It was also used to assist with manuscript drafting, mathematical consistency checks and language editing. All AI-generated suggestions were subject to critical evaluation by the authors, who retain full responsibility for the mathematical derivations, numerical results, physical interpretations and scientific content of the manuscript.

\appendix

\section{Construction of the channel Green function}
\label{app:green}

The scalar Green function provides a compact way to construct all the velocity kernels while satisfying the plate conditions exactly. We use the Fourier convention
\begin{equation}
 \widehat f(k,y)=\int_{-\infty}^{\infty}f(x,y)e^{-ikx}\,\mathrm dx,
 \qquad
 f(x,y)=\frac{1}{2\pi}\int_{-\infty}^{\infty}
 \widehat f(k,y)e^{ikx}\,\mathrm dk.
 \label{app:fourier}
\end{equation}
The streamfunction convention is $u=\partial_y\psi$ and
$v=-\partial_x\psi$. With $q=|k|$, the Green function satisfies
\begin{equation}
 (\partial_y^2-q^2)^2G_q(y,\eta)=\delta(y-\eta),
 \qquad G_q=\partial_yG_q=0\quad\hbox{at }y=0,H.
 \label{app:green-problem}
\end{equation}
Here $y$ and $\eta$ are the target and source heights, respectively.

\subsection{Nonzero wavenumbers}

For $q>0$, introduce $\kappa=qH$ and the two-component columns
\begin{equation}
 \boldsymbol\ell(z)=
 \begin{pmatrix}\sinh z-z\cosh z\\z\sinh z\end{pmatrix},
 \qquad
 \boldsymbol v(z)=\boldsymbol\ell(\kappa-z).
 \label{app:green-bases}
\end{equation}
The first basis satisfies the lower-plate conditions and the second the
upper-plate conditions. The remaining conditions are imposed at the source:
\begin{equation}
 [G_q]=[\partial_yG_q]=[\partial_y^2G_q]=0,
 \qquad [\partial_y^3G_q]=1,
 \label{app:green-jumps}
\end{equation}
where $[f]=f(\eta^+)-f(\eta^-)$. The last relation follows by integrating
\eqref{app:green-problem} across the source. Solving these four matching
conditions gives
\begin{equation}
 G_q(y,\eta)=q^{-3}
 \begin{cases}
 \boldsymbol\ell(qy)^{\mathsf T}\boldsymbol M_\kappa
                   \boldsymbol v(q\eta),&y\leq\eta,\\
 \boldsymbol v(qy)^{\mathsf T}\boldsymbol M_\kappa
                   \boldsymbol\ell(q\eta),&y\geq\eta,
 \end{cases}
 \label{app:green-explicit}
\end{equation}
with
\begin{equation}
 \boldsymbol M_\kappa=
 \frac{1}{2(\sinh^2\kappa-\kappa^2)}
 \begin{pmatrix}
 \sinh\kappa+\kappa\cosh\kappa&\kappa\sinh\kappa\\
 \kappa\sinh\kappa&\kappa\cosh\kappa-\sinh\kappa
 \end{pmatrix}.
 \label{app:green-coupling}
\end{equation}
The matrix is symmetric, giving $G_q(y,\eta)=G_q(\eta,y)$.
Alternatively, this symmetry follows by integrating the self-adjoint
equation \eqref{app:green-problem} by parts; all plate terms vanish.
All derivatives used below are analytic derivatives of the appropriate
branch in \eqref{app:green-explicit}. In particular, differentiating $m$
times in $y$ and $n$ times in $\eta$ introduces a factor $q^{m+n-3}$.

\subsection{Zero mode and small-wavenumber expansion}

Direct evaluation of \eqref{app:green-explicit} suffers cancellation as
$qH\to0$. Let $Y=y/H$, $E=\eta/H$ and write
\begin{equation}
 G_q(y,\eta)=H^3\{P_0(Y,E)+(qH)^2P_2(Y,E)+O((qH)^4)\}.
 \label{app:small-q-green}
\end{equation}
Substitution into \eqref{app:green-problem} gives
\begin{equation}
 \partial_Y^4P_0=\delta(Y-E),\qquad
 \partial_Y^4P_2=2\partial_Y^2P_0,
 \label{app:polynomial-problems}
\end{equation}
with $P_j=\partial_YP_j=0$ at $Y=0,1$. For $Y\leq E$,
\begin{equation}
 P_0(Y,E)=\frac{Y^2E}{2}-Y^2E^2+\frac{Y^2E^3}{2}
          -\frac{Y^3}{6}+\frac{Y^3E^2}{2}-\frac{Y^3E^3}{3},
 \label{app:P0}
\end{equation}
and
\begin{equation}
 \begin{aligned}
 P_2(Y,E)={}&Y^2\left(-\frac{E^2}{15}+\frac{11E^3}{60}
                            -\frac{E^4}{6}+\frac{E^5}{20}\right)\\
 &+Y^3\left(\frac{E^2}{60}-\frac{E^3}{15}
                            +\frac{E^4}{12}-\frac{E^5}{30}\right)\\
 &+Y^4\left(\frac{E}{12}-\frac{E^2}{6}+\frac{E^3}{12}\right)
  +Y^5\left(-\frac{1}{60}+\frac{E^2}{20}-\frac{E^3}{30}\right).
 \end{aligned}
 \label{app:P2}
\end{equation}
For $Y>E$, use $P_j(Y,E)=P_j(E,Y)$. Derivatives are obtained by
differentiating the polynomials on each branch. These expressions also
define the exact zero mode $G_0=H^3P_0$.

\section{Double-layer kernel and completion}
\label{app:double-layer}

\subsection{Stokeslet, rotlet and reciprocal traction}

A point force $\boldsymbol f$ at $\boldsymbol x'=(\xi,\eta)$ enters
the momentum equation as $+\boldsymbol f\delta(\boldsymbol x-\boldsymbol x')$.
For a source at $\xi=0$, the transformed streamfunction and velocity are
\begin{equation}
 \widehat\psi=\frac{1}{\mu}(f_xG_\eta+ikf_yG),\qquad
 \widehat{\boldsymbol S}(k;y,\eta)=\frac{1}{\mu}
 \begin{pmatrix}G_{y\eta}&ikG_y\\-ikG_\eta&k^2G\end{pmatrix},
 \label{app:stokeslet}
\end{equation}
where $G=G_{|k|}(y,\eta)$ and subscripts denote partial derivatives.
The factor $e^{-ik\xi}$ restores the source position. A positive unit
torque is represented by the antisymmetric force dipole
\begin{equation}
 \boldsymbol r(\boldsymbol x,\boldsymbol x')=
 \frac12\left(\partial_\xi\boldsymbol S_{\cdot 2}
                    -\partial_\eta\boldsymbol S_{\cdot 1}\right),
 \qquad
 \widehat{\boldsymbol r}=
 \frac{1}{2\mu}
 \begin{pmatrix}k^2G_y-G_{y\eta\eta}\\
 ik(G_{\eta\eta}-k^2G)\end{pmatrix}.
 \label{app:rotlet}
\end{equation}
The factor $1/2$ gives unit torque, rather than the sum of two unit couples.

To define the double layer, consider the auxiliary flow due to a unit
force in direction $i$ applied at the target $\boldsymbol x$. Evaluate
its stress $\boldsymbol\sigma^{(i)}$ at the source $\boldsymbol x'$.
The contracted kernel is
\begin{equation}
 K_{ij}(\boldsymbol x,\boldsymbol x')=
 -\sigma^{(i)}_{j\ell}(\boldsymbol x')n_\ell(\boldsymbol x'),
 \label{app:reciprocal-traction}
\end{equation}
with summation over $\ell$. This definition includes both the pressure
and viscous parts of the auxiliary stress.

For direct implementation, the following formulas give every component.
For $k=q>0$, define
\begin{equation}
\begin{aligned}
 \Pi_1&=\frac{G_{y\eta\eta\eta}-q^2G_{y\eta}}{q}, &
 \Pi_2&=G_{\eta\eta\eta}-q^2G_\eta,\\
 A&=G_{y\eta\eta}+q^2G_y, &
 B&=G_{\eta\eta}+q^2G.
\end{aligned}
 \label{app:traction-abbreviations}
\end{equation}
Away from the source height,
\begin{align}
 \operatorname{Re}\widehat{\boldsymbol K}
 &=\begin{pmatrix}
 -n_yA&-n_xA\\
 n_x(\Pi_2+2q^2G_\eta)&n_y(\Pi_2-2q^2G_\eta)
 \end{pmatrix},\label{app:stresslet-real}\\
 \operatorname{Im}\widehat{\boldsymbol K}
 &=\begin{pmatrix}
 n_x(\Pi_1+2qG_{y\eta})&n_y(\Pi_1-2qG_{y\eta})\\
 qn_yB&qn_xB
 \end{pmatrix}.
 \label{app:stresslet-imag}
\end{align}
The viscosity cancels between the Stokeslet and its stress, so that
$\boldsymbol K$ is independent of $\mu$. Negative wavenumbers follow
by complex conjugation. Equal-height evaluation and Fourier inversion
are described in appendix~\ref{app:numerics}.

\subsection{Boundary limit and the three density modes}

Split the kernel into its free-space part and a smooth wall correction,
$\boldsymbol K=\boldsymbol K^{(0)}+\boldsymbol K^{(w)}$. With
$\boldsymbol\rho=\boldsymbol x-\boldsymbol x'$, the first is
\begin{equation}
 \boldsymbol K^{(0)}(\boldsymbol x,\boldsymbol x')=
 -\frac{\boldsymbol\rho\boldsymbol\cdot\boldsymbol n'}
        {\pi|\boldsymbol\rho|^4}
        \boldsymbol\rho\boldsymbol\rho^{\mathsf T}.
 \label{app:free-stresslet}
\end{equation}
Only this local singularity contributes to the jump. Integrating the
locally flat-boundary kernel against a constant density gives the
fluid-side contribution $-\boldsymbol\phi/2$; subtracting that constant
density leaves the principal-value integral. Thus
\begin{equation}
 \left.\boldsymbol u^D\right|_{\Gamma,\mathrm{fluid}}
 =-\frac12\boldsymbol\phi+
   \operatorname{p.v.}\!\int_\Gamma
       \boldsymbol K(\boldsymbol x_b,\boldsymbol x')
       \boldsymbol\phi(\boldsymbol x')\,\mathrm ds'.
 \label{app:jump}
\end{equation}

The densities $\boldsymbol e_x$, $\boldsymbol e_y$ and
$\boldsymbol e_z\times(\boldsymbol x'-\boldsymbol x_c)$ generate zero
double-layer velocity outside the cylinder. This follows by applying
the reciprocal identity inside the cylinder to a rigid velocity field:
its strain and stress may be taken to vanish, and the auxiliary
Stokeslet has no singularity in that interior domain. These three
densities therefore lie in the nullspace of the fluid-side boundary
operator. They are removed by
\begin{equation}
 \int_\Gamma\boldsymbol\phi\,\mathrm ds=\boldsymbol0,
 \qquad
 \int_\Gamma\boldsymbol t\boldsymbol\cdot\boldsymbol\phi\,\mathrm ds=0.
 \label{app:density-constraints}
\end{equation}
For a circle, the rotational rigid velocity is $a\boldsymbol t$;
the second relation is therefore precisely the rotational orthogonality
condition, up to a nonzero constant factor.

\subsection{Completion strengths and hydrodynamic loads}

The double layer has zero resultant force and torque: its elementary
singularity is a stresslet, with neither a force monopole nor an
antisymmetric couple. Equivalently, continuity of its traction and
force and moment balance for its source-free interior field give these
zero resultants. The completed field is
\begin{equation}
 \boldsymbol u=\boldsymbol u^D+
 \boldsymbol S(\boldsymbol x,\boldsymbol x_c)\boldsymbol f^c
 +\boldsymbol r(\boldsymbol x,\boldsymbol x_c)\tau^c.
 \label{app:completed-field}
\end{equation}
Integrating the singular momentum equation over a contour enclosing the
centre gives the force resultant $-\boldsymbol f^c$.
Angular-momentum balance similarly gives $- \tau^c$; the
Stokeslet at the centre has zero moment about that centre. The smooth
wall corrections change neither resultant. Consequently,
\begin{equation}
 \boldsymbol F=-\boldsymbol f^c,\qquad
 T=-\tau^c.
 \label{app:completion-loads}
\end{equation}
The constraints \eqref{app:density-constraints} fix the auxiliary
density, while these completion strengths retain the physical loads.
In particular, the rotational unknown is $\tau^c/a$ when
the third completion column is $a\boldsymbol r$; there are still only
two density components at each boundary point.

For homogeneous boundary data, the exterior energy identity first gives
zero velocity and zero loads. Equation~\eqref{app:completion-loads}
then gives zero completion strengths. The remaining double-layer
density is a rigid-motion null mode, which is removed by
\eqref{app:density-constraints}. This explains uniqueness of the
completed system without interpreting its density as a traction.

\section{Numerical evaluation and quadrature}
\label{app:numerics}

\subsection{Boundary nodes and diagonal values}

Let $s_j=2\pi j/N$, $j=0,\ldots,N-1$, be uniformly spaced auxiliary
angles. A smooth map concentrates nodes near the closest wall:
\begin{equation}
 \theta(s)=s-\alpha\sin(s-\theta_g),\qquad
 \theta'(s)=1-\alpha\cos(s-\theta_g),\qquad 0\leq\alpha<1.
 \label{app:mesh-map}
\end{equation}
Here $\theta_g=-\pi/2$ or $\pi/2$ selects the lower or upper gap.
The nodes and arclength weights are
\begin{equation}
 \boldsymbol x_j=(a\cos\theta_j,y_c+a\sin\theta_j),\qquad
 w_j=\frac{2\pi a}{N}\theta'(s_j).
 \label{app:mesh-weights}
\end{equation}
The case $\alpha=0$ recovers uniform angular spacing. The same weights
are used in both the boundary equation and the density constraints.

The free-space diagonal is obtained analytically. For a target displaced
by an angle $\delta$ from the source,
$\boldsymbol\rho=a\boldsymbol t\delta-a\boldsymbol n\delta^2/2
+O(\delta^3)$. Substitution into \eqref{app:free-stresslet} gives
\begin{equation}
 \boldsymbol K^{(0)}_{jj}=
 \frac{\boldsymbol t_j\boldsymbol t_j^{\mathsf T}}{2\pi a}.
 \label{app:free-diagonal}
\end{equation}
The smooth wall correction is added to this value. The jump term
$-\boldsymbol I/2$ in \eqref{app:jump} remains separate: it is not
included in the diagonal kernel value.

\subsection{Fourier inversion and the small-wavenumber pole}

For any kernel with conjugate symmetry, writing
$\widehat{\boldsymbol K}=\boldsymbol K_R+i\boldsymbol K_I$ gives
\begin{equation}
 \boldsymbol K(X;y,\eta)=\frac{1}{\pi}\int_0^\infty
 \{\boldsymbol K_R\cos(qX)-\boldsymbol K_I\sin(qX)\}\,\mathrm dq,
 \qquad X=x-\xi.
 \label{app:positive-inversion}
\end{equation}
For the double layer, the known transform of
\eqref{app:free-stresslet} is subtracted before quadrature, and its
real-space value is added back. This isolates the local singularity
from the Fourier integration.

There is also a low-wavenumber contribution. From
\eqref{app:P0} one obtains
\begin{equation}
 G_{0,y\eta\eta\eta}=\frac{6y(H-y)}{H^3},\qquad
 \widehat{\boldsymbol K}=
 \frac{i}{k}\boldsymbol A_{-1}(y,\boldsymbol n')
 +\boldsymbol A_0(y,\eta,\boldsymbol n')+O(k),
 \label{app:kernel-pole}
\end{equation}
where
\begin{equation}
 \boldsymbol A_{-1}=
 \frac{6y(H-y)}{H^3}
 \begin{pmatrix}n_x'&n_y'\\0&0\end{pmatrix}.
 \label{app:pole-coefficient}
\end{equation}
The pole is interpreted in the principal-value Fourier sense. For
numerical integration, subtract
$i\boldsymbol A_{-1}e^{-(k/k_c)^2}/k$ and add its exact inverse,
\begin{equation}
 -\frac12\boldsymbol A_{-1}
       \operatorname{erf}\!\left(\frac{k_cX}{2}\right).
 \label{app:pole-inverse}
\end{equation}
The Gaussian preserves the decay at large wavenumbers and leaves a
regular integrand near the origin.

The integrated pole is proportional to
$\int_\Gamma\boldsymbol n\boldsymbol\cdot\boldsymbol\phi\,\mathrm ds$.
For the solution of the boundary equation this integral is zero. To
see this, the double-layer traces satisfy
$\boldsymbol u^D_{\mathrm{fluid}}-\boldsymbol u^D_{\mathrm{interior}}
=-\boldsymbol\phi$. The interior field is divergence-free, and the
Stokeslet and rotlet completions have zero net volume source. Integrating
the normal component of the boundary equation therefore gives
\begin{equation}
 \int_\Gamma\boldsymbol u_b\boldsymbol\cdot\boldsymbol n\,\mathrm ds
 =-\int_\Gamma\boldsymbol\phi\boldsymbol\cdot\boldsymbol n\,\mathrm ds=0.
 \label{app:flux-compatibility}
\end{equation}
This compatibility follows from the prescribed rigid velocity; it is
not a fourth density constraint. It also holds for the pump disturbance
data, since the imposed Poiseuille field has zero net flux through the
closed cylinder contour. The pole must nevertheless be retained in each
kernel evaluation, including its finite source-position contribution
to the pressure functional in appendix~\ref{app:pressure}.

\subsection{Large-wavenumber tails}

The numerical Fourier integral is truncated at $q_{\max}$. In a narrow
gap, the nearest wall produces a slowly decaying image contribution
that is integrated analytically beyond this cutoff. If $b$ and $c$ are
the target and source distances from that wall, its scalar Green
correction is
\begin{equation}
 G_q^{\mathrm{img}}(b,c)=
 -\frac{1+q(b+c)+2q^2bc}{4q^3}e^{-q(b+c)}.
 \label{app:image-green}
\end{equation}
For the lower wall $(b,c)=(y,\eta)$; for the upper wall
$(b,c)=(H-y,H-\eta)$. All derivatives are taken with respect to the
original coordinates, so that the upper-wall reflection changes the
sign of each odd vertical derivative. Substituting this scalar
correction into \eqref{app:stokeslet}, \eqref{app:rotlet} and
\eqref{app:stresslet-real}--\eqref{app:stresslet-imag} determines the
corresponding kernel tails.

These tails reduce to moments
\begin{equation}
 J_n=\int_{q_{\max}}^\infty q^n e^{-zq}\,\mathrm dq,
 \qquad z=b+c-iX,
 \label{app:tail-moments}
\end{equation}
with
\begin{equation}
 \begin{aligned}
 J_{-1}&=E_1(q_{\max}z),&
 J_0&=\frac{e^{-q_{\max}z}}{z},\\
 J_1&=e^{-q_{\max}z}
      \left(\frac{q_{\max}}{z}+\frac{1}{z^2}\right),&
 J_2&=e^{-q_{\max}z}
      \left(\frac{q_{\max}^2}{z}+\frac{2q_{\max}}{z^2}
                                      +\frac{2}{z^3}\right).
 \end{aligned}
 \label{app:tail-explicit}
\end{equation}
Here $E_1$ is the exponential integral; cosine and sine moments are the
real and imaginary parts. For the completion fields, the free-space
tail is also retained. It follows from the scalar kernel
\begin{equation}
 G_q^{(0)}(y,\eta)=
 \frac{1+q|y-\eta|}{4q^3}e^{-q|y-\eta|},
 \label{app:free-green}
\end{equation}
by the same differentiation and integration. The double-layer
free-space contribution already includes its entire Fourier tail,
because it was added in real space. More distant wall reflections are
left in the truncated remainder; their omission is controlled by
checking convergence with $q_{\max}$.

\subsection{Evaluation details and convergence checks}

The implementation evaluates lengths in units of $H$. It uses the
polynomial expansion for $qH<10^{-3}$, the Gaussian cutoff $k_cH=4$,
and composite 16-point Gauss--Legendre quadrature in wavenumber. The
initial panel edges in $qH$ are
$0,10^{-3},0.05,0.1,0.25,0.5,1,2$; subsequent panels have width $2$
up to $20$ and width $4$ thereafter, with the last panel ending at the
selected cutoff. Small arguments in the hyperbolic bases and coupling
coefficients are evaluated by series. The cutoff and boundary-node
count are numerical parameters, rather than physical truncations of
the channel.

Heights differing only by round-off are treated as equal. At equal
target and source heights, analytic derivatives are evaluated by
averaging their two one-sided branch values, consistently with the
free-space subtraction; in particular, the free-space factor
$\operatorname{sgn}(y-\eta)$ is set to zero. The analytic free-space
completion tails are evaluated by taking the equal-height limit
after Fourier integration. No finite-difference derivatives of
the Green function are needed.

Boundary accuracy is assessed at auxiliary midpoint angles
$s_j+\pi/N$, in addition to the collocation nodes. The density is
interpolated periodically in $s$, and the reconstructed boundary
velocity includes the term $-\boldsymbol\phi/2$. For a graded mesh,
Fourier interpolation in the nonuniform physical angle $\theta$ would
not give the same interpolant. Further checks are the density
constraints, \eqref{app:flux-compatibility}, reciprocal symmetry of
the resistance matrix, and agreement of pressure jumps evaluated
above and below the cylinder. The node count and Fourier cutoff are
varied separately. With the hydrodynamic-load convention of the main
text, $-\widehat{\boldsymbol R}$, rather than
$\widehat{\boldsymbol R}$, is positive definite.

\section{Pressure jump, reciprocity and the pump calculation}
\label{app:pressure}

\subsection{Pressure from the velocity zero mode}

Choose a horizontal line below or above the entire cylinder so that
it does not intersect a singularity or the solid. The transformed
streamwise momentum equation is
\begin{equation}
 ik\widehat p=\mu(\partial_y^2-k^2)\widehat u.
 \label{app:pressure-momentum}
\end{equation}
For a zero-through-flow solution, $p$ tends to constants $p_-$ and
$p_+$ at the two ends. In the distributional Fourier sense,
$\lim_{k\to0}ik\widehat p=p_+-p_-$. Therefore
\begin{equation}
 \Delta p=p_--p_+=-\mu\partial_y^2\widehat u(0,y).
 \label{app:pressure-zero-mode}
\end{equation}
This formula is independent of the arbitrary common pressure constant.

The source-position phase is essential when evaluating the zero mode.
Using \eqref{app:kernel-pole},
\begin{equation}
 e^{-ik\xi}\widehat{\boldsymbol K}
 =\frac{i}{k}\boldsymbol A_{-1}
  +\xi\boldsymbol A_{-1}+\boldsymbol A_0+O(k).
 \label{app:source-phase}
\end{equation}
Although the first term cancels after integration by
\eqref{app:flux-compatibility}, the finite term
$\xi\boldsymbol A_{-1}$ does not. Define
$m(\boldsymbol x')=n_y'\phi_x+n_x'\phi_y$ and
$g(\boldsymbol x')=\boldsymbol n'\boldsymbol\cdot\boldsymbol\phi$.
The regular first-row double-layer contribution is
$-G_{0,y\eta\eta}m$. Combining it with the source phase and the
completion fields gives
\begin{equation}
 \begin{aligned}
 \Delta p={}&\mu\int_\Gamma
       G_{0,yyy\eta\eta}(y,\eta)m(\boldsymbol x')\,\mathrm ds'
       +\frac{12\mu}{H^3}\int_\Gamma\xi g(\boldsymbol x')\,\mathrm ds'\\
 &-G_{0,yyy\eta}(y,y_c)f_x^c
       +\frac12G_{0,yyy\eta\eta}(y,y_c)\tau^c.
 \end{aligned}
 \label{app:pressure-functional}
\end{equation}
The vertical Stokeslet makes no contribution to this zero mode.

For a target line below the cylinder, differentiation of
\eqref{app:P0} gives
\begin{equation}
 G_{0,yyy\eta}(y,\eta)=\frac{6\eta(H-\eta)}{H^3},\qquad
 G_{0,yyy\eta\eta}(y,\eta)=\frac{6(H-2\eta)}{H^3}.
 \label{app:pressure-derivatives}
\end{equation}
Thus a particularly simple dimensional form is
\begin{equation}
 \begin{aligned}
 \Delta p={}&\frac{6\mu}{H^3}\int_\Gamma
       (H-2\eta)(n_y'\phi_x+n_x'\phi_y)\,\mathrm ds'
       +\frac{12\mu}{H^3}\int_\Gamma
       \xi\,\boldsymbol n'\boldsymbol\cdot\boldsymbol\phi\,\mathrm ds'\\
 &-\frac{6y_c(H-y_c)}{H^3}f_x^c
       +\frac{3(H-2y_c)}{H^3}\tau^c.
 \end{aligned}
 \label{app:pressure-explicit}
\end{equation}
The upper branch gives the same result for the completed solution.
In the discretisation, both integrals use the boundary weights
\eqref{app:mesh-weights}. Applying the functional to the three rigid
motions gives $P^U$, $P^V$ and $P^\Omega$ through
\begin{equation}
 \Delta p=\frac{\mu}{H}
       (P^U U+P^V V+P^\Omega a\Omega).
 \label{app:pressure-coefficients}
\end{equation}

\subsection{Imposed flux and calculation of the additional resistance}

For a rotating cylinder with fixed centre, an imposed flux $Q$ is
introduced by separating the fully developed empty-channel solution:
\begin{equation}
 \boldsymbol u_P(y)=\frac{6Q}{H^3}y(H-y)\boldsymbol e_x,
 \qquad p_P(x)=-\frac{12\mu Q}{H^3}x,
 \qquad \boldsymbol u=\boldsymbol u_P+\boldsymbol u',\quad p=p_P+p'.
 \label{app:pump-decomposition}
\end{equation}
The disturbance has zero wall velocity, decays at both ends and obeys
\begin{equation}
 \left.\boldsymbol u'\right|_\Gamma=
 a\Omega\boldsymbol t-\boldsymbol u_P(y_b).
 \label{app:pump-boundary}
\end{equation}
Thus the original infinite-channel kernels and augmented matrix apply
unchanged to the disturbance. The linearly varying $p_P$ is known
analytically and is never evaluated as a difference between pressures
at infinity. The finite excess jump is
$\Delta p_{\mathrm{ex}}=p'_- -p'_+$.

To determine $C$, set $\Omega=0$ and solve the fixed-cylinder problem
with boundary data $-\boldsymbol u_P$. The resulting density and
completion strengths are substituted into
\eqref{app:pressure-functional}, giving
\begin{equation}
 C=\frac{H^2}{\mu Q}\Delta p_{\mathrm{ex}}(0,Q).
 \label{app:pump-C}
\end{equation}
This is the additional hydraulic resistance due to the cylinder; it
requires a separate through-flow right-hand side and is not determined
by the three rigid-motion resistance columns alone. With a reference
speed $U_0$ and $Q=HU_0$, the additional nodal right-hand side is simply
\begin{equation}
 \boldsymbol b_j^{(Q)}=-6U_0Y_j(1-Y_j)\boldsymbol e_x,
 \qquad Y_j=y_j/H.
 \label{app:pump-fourth-rhs}
\end{equation}
The three constraint entries remain zero. The Poiseuille background
is smooth and source-free throughout the disk occupied by the cylinder,
so its stress has zero integrated force and torque on that closed
contour. Consequently the disturbance completion strengths also give
the total cylinder loads in the imposed-flux problem.

\subsection{Reciprocity and the pressure--flow characteristic}

Apply the reciprocal theorem to solution A, a rotating cylinder at
zero flux, and solution B, a stationary cylinder in imposed flux $Q$.
The plate contributions vanish. The cylinder contribution on the A
side is $-\Omega T^{(Q)}$, while the remote-section contribution on
the B side is $\Delta p^{(\Omega)}Q$. The other end contribution
vanishes because the velocity of A decays even though the pressure
of B grows linearly. Hence
\begin{equation}
 -\Omega T^{(Q)}=\Delta p^{(\Omega)}Q,
 \qquad
 T^{(Q)}=-\frac{\mu aQ}{H}P^\Omega.
 \label{app:pump-reciprocity}
\end{equation}
Here $T^{(Q)}$ is the hydrodynamic torque, consistently with the main
text. Superposition then gives
\begin{align}
 \Delta p_{\mathrm{ex}}
 &=\frac{\mu a\Omega}{H}P^\Omega+\frac{\mu Q}{H^2}C,
 \label{app:pump-characteristic}\\
 T&=\mu a^2\widehat R^{T\Omega}\Omega
                         -\frac{\mu aQ}{H}P^\Omega.
 \label{app:pump-torque}
\end{align}

For comparison with pressure measurements in a finite channel, choose
sections $x_-$ and $x_+$ in the developed regions on either side of the
cylinder, with $L=x_+-x_-$. Restoring the empty-channel pressure drop
gives
\begin{equation}
 \Delta p_L=p(x_-)-p(x_+)
 \simeq\frac{\mu a\Omega}{H}P^\Omega
       +\frac{\mu Q}{H^2}\left(C+\frac{12L}{H}\right).
 \label{app:finite-channel-pressure}
\end{equation}
At equal section pressures the corresponding free-delivery flux is
\begin{equation}
 Q_{\mathrm{free}}=-\frac{a\Omega H P^\Omega}{C+12L/H}.
 \label{app:free-delivery}
\end{equation}
The delivered pressure rise in the positive $x$ direction is
$-\Delta p_L$. Equation~\eqref{app:finite-channel-pressure} uses the
infinite-channel disturbance and assumes that the measurement sections
and physical end conditions are sufficiently remote. It does not
include entrance, reservoir or nearby-end effects. In particular,
$\Delta p_{\mathrm{ex}}=0$ removes only the excess pressure difference;
it is not the equal-pressure condition for a finite channel.

\bibliographystyle{plainnat}
\bibliography{References}

@article{Faxen1946,
  author  = {Fax\'en, O. H.},
  title   = {Forces exerted on a rigid cylinder in a viscous fluid between two parallel fixed planes},
  journal = {Proc. R. Swed. Inst. Eng. Res.},
  volume  = {187},
  pages   = {1--13},
  year    = {1946}
}

@article{Takaisi1955,
  author  = {Takaisi, Y.},
  title   = {The drag on a circular cylinder moving with low speeds in a viscous liquid between two parallel walls},
  journal = {J. Phys. Soc. Jpn},
  volume  = {10},
  pages   = {685--693},
  year    = {1955}
}

@article{Harper1967,
  author  = {Harper, E. Y. and Chang, I.-D.},
  title   = {Drag on a cylinder between parallel walls in {S}tokes' flow},
  journal = {Phys. Fluids},
  volume  = {10},
  pages   = {83--88},
  year    = {1967}
}

@article{Jeffrey1981,
  author  = {Jeffrey, D. J. and Onishi, Y.},
  title   = {The slow motion of a cylinder next to a plane wall},
  journal = {Q. J. Mech. Appl. Maths},
  volume  = {34},
  pages   = {129--137},
  year    = {1981}
}

@article{Dvinsky1987a,
  author  = {Dvinsky, A. S. and Popel, A. S.},
  title   = {Motion of a rigid cylinder between parallel plates in {S}tokes flow. {P}art 1: {M}otion in a quiescent fluid and sedimentation},
  journal = {Comput. Fluids},
  volume  = {15},
  pages   = {391--404},
  year    = {1987}
}

@article{Dvinsky1987b,
  author  = {Dvinsky, A. S. and Popel, A. S.},
  title   = {Motion of a rigid cylinder between parallel plates in {S}tokes flow. {P}art 2: {P}oiseuille and {C}ouette flow},
  journal = {Comput. Fluids},
  volume  = {15},
  pages   = {405--419},
  year    = {1987}
}

@article{Pozrikidis1987,
  author  = {Pozrikidis, C.},
  title   = {Creeping flow in two-dimensional channels},
  journal = {J. Fluid Mech.},
  volume  = {180},
  pages   = {495--514},
  year    = {1987}
}

@article{Davis1993,
  author  = {Davis, A. M. J.},
  title   = {Periodic blocking in parallel shear or channel flow at low {R}eynolds number},
  journal = {Phys. Fluids A},
  volume  = {5},
  pages   = {800--809},
  year    = {1993}
}

@article{Crowdy2013,
  author  = {Crowdy, D. G. and Davis, A. M. J.},
  title   = {Stokes flow singularities in a two-dimensional channel: a novel transform approach with application to microswimming},
  journal = {Proc. R. Soc. Lond. A},
  volume  = {469},
  pages   = {20130198},
  year    = {2013}
}

@article{Sen1996,
  author  = {Sen, M. and Wajerski, D. and Gad-el-Hak, M.},
  title   = {A novel pump for {MEMS} applications},
  journal = {Trans. ASME J. Fluids Engng},
  volume  = {118},
  pages   = {624--627},
  year    = {1996}
}

@article{Day2000,
  author  = {Day, R. F. and Stone, H. A.},
  title   = {Lubrication analysis and boundary integral simulations of a viscous micropump},
  journal = {J. Fluid Mech.},
  volume  = {416},
  pages   = {197--216},
  year    = {2000}
}

@article{Teng2022,
  author  = {Teng, Jian and Rallabandi, Bhargav and Stone, Howard A. and Ault, Jesse T.},
  title   = {Coupling of translation and rotation in the motion of finite-length rods near solid boundaries},
  journal = {J. Fluid Mech.},
  volume  = {938},
  pages   = {A30},
  year    = {2022},
  doi     = {10.1017/jfm.2022.177}
}

@article{Champmartin2007a,
  author  = {Champmartin, S. and Ambari, A.},
  title   = {Kinematics of a symmetrically confined cylindrical particle in a `{S}tokes-type' regime},
  journal = {Phys. Fluids},
  volume  = {19},
  pages   = {073303},
  year    = {2007}
}

@article{Champmartin2007b,
  author  = {Champmartin, S. and Ambari, A. and Roussel, N.},
  title   = {Flow around a confined rotating cylinder at small {R}eynolds number},
  journal = {Phys. Fluids},
  volume  = {19},
  pages   = {103101},
  year    = {2007}
}

@article{Yang2011,
  author  = {Yang, J. and Huber, G. and Wolgemuth, C. W.},
  title   = {Forces and torques on rotating spirochete flagella},
  journal = {Phys. Rev. Lett.},
  volume  = {107},
  pages   = {268101},
  year    = {2011}
}

@article{Yang2013,
  author  = {Yang, J. and Wolgemuth, C. W. and Huber, G.},
  title   = {Force and torque on a cylinder rotating in a narrow gap at low {R}eynolds number: scaling and lubrication analyses},
  journal = {Phys. Fluids},
  volume  = {25},
  pages   = {051901},
  year    = {2013}
}

@article{Jeong2014a,
  author  = {Jeong, J.-T. and Yoon, S.-H.},
  title   = {Two-dimensional {S}tokes flow around a circular cylinder in a microchannel},
  journal = {J. Mech. Sci. Technol.},
  volume  = {28},
  pages   = {573--579},
  year    = {2014}
}

@article{Jeong2014b,
  author  = {Jeong, J.-T. and Jang, C.-S.},
  title   = {Slow motion of a circular cylinder in a plane {P}oiseuille flow in a microchannel},
  journal = {Phys. Fluids},
  volume  = {26},
  pages   = {123104},
  year    = {2014}
}

@article{Power1987,
  author  = {Power, H. and Miranda, G.},
  title   = {Second kind integral equation formulation of {S}tokes' flows past a particle of arbitrary shape},
  journal = {SIAM J. Appl. Maths},
  volume  = {47},
  pages   = {689--698},
  year    = {1987}
}

@article{Karrila1989,
  author  = {Karrila, S. J. and Kim, S.},
  title   = {Integral equations of the second kind for {S}tokes flow: direct solution for physical variables and removal of inherent accuracy limitations},
  journal = {Chem. Engng Commun.},
  volume  = {82},
  pages   = {123--161},
  year    = {1989}
}

@article{PhanThien1992,
  author  = {Phan-Thien, N. and Tullock, D. and Kim, S.},
  title   = {Completed double layer in half-space: a boundary element method},
  journal = {Comput. Mech.},
  volume  = {9},
  pages   = {121--135},
  year    = {1992}
}

@article{Power1993a,
  author  = {Power, H.},
  title   = {The completed double layer boundary integral equation method for two-dimensional {S}tokes flow},
  journal = {IMA J. Appl. Maths},
  volume  = {51},
  pages   = {123--145},
  year    = {1993}
}

@article{Power1993b,
  author  = {Power, H. and Febres de Power, B.},
  title   = {Second-kind integral equation formulation for the slow motion of a particle of arbitrary shape near a plane wall in a viscous fluid},
  journal = {SIAM J. Appl. Maths},
  volume  = {53},
  pages   = {60--70},
  year    = {1993}
}

@article{Kohr2005,
  author  = {Kohr, M.},
  title   = {A second-kind integral equation method for {S}tokes flow past smooth obstacles in a channel},
  journal = {Mathematica (Cluj)},
  volume  = {47 (70)},
  pages   = {165--178},
  year    = {2005}
}

@article{Topuz2021,
  author  = {Topuz, A. and Baranoglu, B. and \c{C}etin, B.},
  title   = {A multi-domain direct boundary element formulation for particulate flow in microchannels},
  journal = {Engng Anal. Bound. Elem.},
  volume  = {132},
  pages   = {221--230},
  year    = {2021}
}

@article{Gumus2025,
  author  = {Gumus, O. C. and Kabacaoglu, G. and \c{C}etin, B.},
  title   = {Isogeometric boundary element formulation to simulate droplets in microchannel confinement},
  journal = {Intl J. Numer. Meth. Heat Fluid Flow},
  volume  = {35},
  pages   = {1965--1982},
  year    = {2025}
}

@article{Karakaya2015,
  author  = {Karakaya, Z. and Baranoglu, B. and \c{C}etin, B. and Yazici, A.},
  title   = {A parallel boundary element formulation for tracking multiple particle trajectories in {S}tokes' flow for microfluidic applications},
  journal = {Comput. Model. Engng Sci.},
  volume  = {104},
  pages   = {227--249},
  year    = {2015}
}

@article{Kohr1998,
  author  = {Kohr, M.},
  title   = {An integral method for two-dimensional {S}tokes flows past a rigid obstacle in the half-plane},
  journal = {Appl. Mech. Engng},
  volume  = {3},
  pages   = {5--24},
  year    = {1998}
}

@article{Kohr2000,
  author  = {Kohr, M.},
  title   = {An indirect boundary integral method for a {S}tokes flow problem},
  journal = {Comput. Meth. Appl. Mech. Engng},
  volume  = {190},
  pages   = {487--497},
  year    = {2000}
}

@article{Bhattacharya2005,
  author  = {Bhattacharya, S. and B{\l}awzdziewicz, J. and Wajnryb, E.},
  title   = {Hydrodynamic interactions of spherical particles in suspensions confined between two planar walls},
  journal = {J. Fluid Mech.},
  volume  = {541},
  pages   = {263--292},
  year    = {2005}
}

@article{DeCourtye1998,
  author  = {DeCourtye, D. and Sen, M. and Gad-el-Hak, M.},
  title   = {Analysis of viscous micropumps and microturbines},
  journal = {Intl J. Comput. Fluid Dyn.},
  volume  = {10},
  pages   = {13--25},
  year    = {1998}
}

@article{Hellou1992,
  author  = {Hellou, M. and Coutanceau, M.},
  title   = {Cellular {S}tokes flow induced by rotation of a cylinder in a closed channel},
  journal = {J. Fluid Mech.},
  volume  = {236},
  pages   = {557--577},
  year    = {1992}
}

@article{Liron1976,
  author  = {Liron, N. and Mochon, S.},
  title   = {Stokes flow for a stokeslet between two parallel flat plates},
  journal = {J. Engng Maths},
  volume  = {10},
  pages   = {287--303},
  year    = {1976}
}

@article{Sharatchandra1997,
  author  = {Sharatchandra, M. C. and Sen, M. and Gad-el-Hak, M.},
  title   = {Navier--{S}tokes simulations of a novel viscous pump},
  journal = {Trans. ASME J. Fluids Engng},
  volume  = {119},
  pages   = {372--382},
  year    = {1997}
}

@article{Staben2003,
  author  = {Staben, M. E. and Zinchenko, A. Z. and Davis, R. H.},
  title   = {Motion of a particle between two parallel plane walls in low-{R}eynolds-number {P}oiseuille flow},
  journal = {Phys. Fluids},
  volume  = {15},
  pages   = {1711--1733},
  year    = {2003}
}

\end{document}